# Improved Model Reduction and Tuning of Fractional Order PI$^\lambda$D$^\mu$ Controllers for Analytical Rule Extraction with Genetic Programming

Saptarshi Das[a,b], Indranil Pan[b], Shantanu Das[c] and Amitava Gupta[a,b]

a) School of Nuclear Studies & Applications (SNSA), Jadavpur University, Salt Lake Campus, LB-8, Sector 3, Kolkata-700098, India. Email: saptarshi@pe.jusl.ac.in, amitg@pe.jusl.ac.in

b) Department of Power Engineering, Jadavpur University, Salt Lake Campus, LB-8, Sector 3, Kolkata-700098, India. Email: indranil.jj@student.iitd.ac.in, indranil@pe.jusl.ac.in

c) Reactor Control Division, Bhabha Atomic Research Centre, Mumbai-4000854, India. Email: shantanu@magnum.barc.gov.in

**Abstract:**
Genetic Algorithm (GA) has been used in this paper for a new approach of sub-optimal model reduction in the Nyquist plane and optimal time domain tuning of PID and fractional order (FO) PI$^\lambda$D$^\mu$ controllers. Simulation studies show that the Nyquist based new model reduction technique outperforms the conventional H$_2$ norm based reduced parameter modeling technique. With the tuned controller parameters and reduced order model parameter data-set, optimum tuning rules have been developed with a test-bench of higher order processes via Genetic Programming (GP). The GP performs a symbolic regression on the reduced process parameters to evolve a tuning rule which provides the best analytical expression to map the data. The tuning rules are developed for a minimum time domain integral performance index described by weighted sum of error index and controller effort. From the reported Pareto optimal front of GP based optimal rule extraction technique a trade-off can be made between the complexity of the tuning formulae and the control performance. The efficacy of the single-gene and multi-gene GP based tuning rules has been compared with original GA based control performance for the PID and PI$^\lambda$D$^\mu$ controllers, handling four different class of representative higher order processes. These rules are very useful for process control engineers as they inherit the power of the GA based tuning methodology, but can be easily calculated without the requirement for running the computationally intensive GA every time. Three dimensional plots of the required variation in PID/FOPID controller parameters with reduced process parameters have been shown as a guideline for the operator. Parametric robustness of the reported GP based tuning rules has also been shown with credible simulation examples.

**Keywords:** Automatic rule generation; fractional order PID controller; Genetic Programming; model reduction; optimal time domain tuning; FOPID tuning rule.

**1. Introduction:**
Empirical rules are classically used to tune PID controllers and are very popular in process control since the advent of PID controllers. These rules are mainly devised from certain design specification in time or frequency domain. O' Dwyer [1] has tabulated several optimal PI/PID controller tuning rules for various types of reduced order processes based on diverse control objectives like set-point tracking, load disturbance rejection etc. The conventional step-response process reaction curve based graphical method to obtain First Order Plus Time Delay (FOPTD) models for unknown processes has been extended by Skogestad in [2] for PID controller tuning. Performance comparison of well established empirical rules like Ziegler-Nichols, refined Ziegler-Nichols (Z-N), Cohen-Coon (C-C), Internal Model Control (IMC), Gain-Phase Margin (GPM) have been studied by Tan *et al.* [3] and Lin *et al.* [4]. Also, Ho *et al.* done a comparative study of integral performance index based optimum parameter settings for PI controllers in [5] and PID controllers in [6]. Impact of choosing different performance indices like Integral of Time Multiplied Absolute Error (ITAE) or Integral of Time Multiplied Squared Error (ITSE) corresponding to set-point tracking and load disturbance rejection on the optimum tuning formula have been studied by Zhuang & Atherton [7]. The idea has been extended in Mann *et al.* [8] considering actuator constraints. Ho *et al.* [9] combined the concept of time domain performance index optimization and gain-phase margin based method to develop improved tuning rules.

It is well known that for the development of tuning formula for an arbitrary higher order process, it needs to be reduced first in a suitable template like FOPTD or Second Order Plus Time Delay (SOPTD) etc, since these rules are basically a mapping between the process and optimum controller parameters. Zhuang & Atherton [7] proposed the tuning rule for PID controllers to handle FOPTD processes, which is rather poor approximation for higher order processes as shown by Astrom & Hagglund [10]. Zhuang & Atherton in [7] used several higher moments of time and error terms in the integral performance index which puts higher penalties for larger error and sluggish response, yielding large control signal which may saturate the actuator. This paper tries to extend the idea for the tuning of PID and FOPID controllers with a customized control objective, comprising of a suitable integral error index and the control signal which can be viewed like a trade-off between the ability of set-point tracking and required controller effort [11]-[12]. The optimum time domain tuning of PID type controllers are attempted with genetic algorithm as studied with similar objectives [11]-[14]. These optimal integral performance indices based tuning methods for PID controllers show nice closed loop behavior in terms of low overshoot and settling-time but the only requirement is that the process model has to be identified accurately. These accurate reduced order model parameters can then be used to find out the controller tuning rules, represented by a nonlinear mapping between the process parameters to controller parameters. Similar nonlinear mapping with Artificial Neural Network (ANN) has been applied in [15] to find out FOPID parameters. But the ANN based method in [15] does not produce analytical expressions unlike the present GP based tuning rule extraction technique which is easy to compute and helpful in automation industry. For higher order process models simple FOPTD reduced order approximations give larger modeling errors which may produce inferior closed loop response with the available controller tuning rules. Hence, an improved sub-optimal model reduction in the Nyquist plane is attempted first to reduce few classes of higher order processes in SOPTD template which is a better approximation than the corresponding FOPTD models [10]. Reduction in SOPTD template for improved frequency domain tuning of PID controllers has been extensively studied by Wang *et al.* [16].

Also, Zhuang & Atherton in [7] developed the optimum tuning formula based on the least-square curve fitting technique with a test data-set of optimum controller parameters with few FOPTD models. Such a chosen structure based linear fitting method indeed reduces the accuracy of the tuning formula which is further enhanced in this paper with a much sophisticated technique i.e. a Genetic Programming based approach. It is well known that application of fractional calculus is getting increasing interest in the research community due its higher capability to model and control physical systems [17]-[20]. Conventional notion of PID controllers in process control has been extended by Podlubny [21] with fractional order $PI^\lambda D^\mu$ controllers having higher degrees of freedom for control as the integro-differential orders along with the proportional and integro-differential gains. Since the advent of $PI^\lambda D^\mu$ controller various methods have been proposed by contemporary researchers for its efficient tuning for process control applications. Detailed survey regarding tuning of fractional order $PI^\lambda D^\mu$ controllers can be found in [22]-[25]. Time and error moment approaches of FOPID tuning have been studied in [26]-[27]. Few contemporary researchers like Valerio & Sa da Costa [28], Chen *et al.* [29], Padula & Visioli [30] have developed analytical tuning rules for FOPID controllers. Valerio & Sa da Costa [28] reported step-response process reaction curve based Ziegler-Nichols type FOPID tuning rules. Chen *et al.* [29] proposed a fractional-MIGO based tuning rule FOPI controllers to handle FOPTD processes. Tuning rules for optimum FOPID controllers with minimum IAE with sensitivity constraint has been developed by Padula & Visioli [30]. Gude & Kahoraho [31]-[32] developed tuning rules for FOPI controllers similar to the Ziegler-Nichols open loop (time domain) and closed loop (frequency domain) method and tested the rules for a wide class of higher order processes. ISE based simple optimal tuning rules have been developed by Bayat [33] for varying level of normalized dead-time. The idea of the present paper is to extract the tuning rules in an optimal fashion via GP with initially GA based sub-optimum reduced-parameter-models and optimum PID/FOPID parameters. The rationale behind using Genetic Programming is the fact that it is based on symbolic regression which searches for not only the optimal parameters within a structure but also the structure itself, representing the optimal PID/FOPID controller tuning rules in our case that ensures low error index and control signal. Preliminary results on this investigation have been reported in [34] with low complexity rules and the idea has been extended here for tuning rules of higher complexity and better control performance.

The rest of the paper is organized as follows. Section 2 discusses about a new sub-optimal model reduction for higher order processes. Section 3 shows the GA based optimal PID/FOPID controller tuning results and GP based analytical tuning rule extraction with the achievable closed loop performances. The paper ends with conclusion in section 4, followed by the references.

## 2. An improved sub-optimal model reduction technique:
### *2.1. New optimization framework for model reduction in Nyquist plane:*

It is well known that model reduction refers to compact representation of process models without loss of its dominant dynamic behaviors. Since, the impulse input persistently excites a process model, hence model reduction should be attempted with impulse response characteristics over commonly used step/ramp response. This approach captures delicates dynamic behaviors of a higher order model and an optimization based search for reduced order model parameters would give a highly accurate low complexity process model. In frequency domain the impulse response characteristics is equivalent to the $H_2$-norm of the model. Xue & Chen [35] proposed a novel method of reducing higher order process models by minimizing the $H_2$-norm of the original higher order $P(s)$ and reduced order process $\tilde{P}(s)$ with an unconstrained optimization process. i.e.

$$J_{2-norm} = \left\| P(s) - \tilde{P}(s) \right\|_2 \tag{1}$$

where, $\left\| \cdot \right\|_2$ denotes the 2-norm of a system which is a measure of the energy of a stable LTI system with an impulse excitation and is given by the following expression:

$$\left\| P(s) \right\|_2 = \sqrt{\frac{1}{2\pi} \int_{-\infty}^{\infty} trace\left[ P(j\omega)\overline{P(j\omega)^T} \right] d\omega} \tag{2}$$

Fractional order model reduction approaches using similar kind of $H_2$-norm based optimization framework has been studied in [26]. Other relevant works include dominant mode based methods [36]. Advancements on the FO model reduction techniques have been illustrated in a detail manner in [37]-[38].

In this paper another optimization framework has been used which minimizes the discrepancy between the frequency responses of the higher order and reduced parameter process model in the complex Nyquist plane. The proposed methodology has been found to produce better accuracy in the model reduction process, since the $H_2$-norm based method, discussed earlier [35] is based on the minimization of the discrepancy in the magnitude of the frequency response only. Whereas, the proposed Nyquist based method minimizes both the discrepancies in the magnitude and phase of the two said systems. The proposed objective function for model reduction is given by (3):

$$J_{nyquist} = w_1 \cdot \left\| \text{Re}\left[ P(j\omega) \right] - \text{Re}\left[ \tilde{P}(j\omega) \right] \right\| + w_2 \cdot \left\| \text{Im}\left[ P(j\omega) \right] - \text{Im}\left[ \tilde{P}(j\omega) \right] \right\| \tag{3}$$

Here, the norm $\left\| \cdot \right\|$ denotes Euclidian length of the vectors. The weights $\{w_1, w_2\}$ are chosen to be equal so as not to emphasize discrepancies either in the real or imaginary part of the transfer function. To evaluate the objective function (3) in each iteration, within an optimization framework, logarithmically spaced 500 frequency points have been taken within the frequency-band of $\omega \in [\omega_l, \omega_h] = [10^{-4}, 10^4] Hz$. Here, the two objective functions (1) and (3) denotes the discrepancies in the $H_2$-norm and the real and imaginary parts of the Nyquist curves corresponding to the higher order process and the reduced order models. The objective functions (1) and (3) are minimized with an unconstrained Genetic Algorithm to obtain the reduced parameter models in a FOPTD (5) as well as SOPTD (6) templates with the corresponding sub-optimal reduced order parameters in Table 1-2 for a test-bench of higher order processes. The model reduction technique has been termed as "sub-optimal" due to the fact that it extracts the apparent delays ($L$) in the higher order models with an equivalent third order Pade approximation:

$$e^{-Ls} \simeq \frac{-L^3 s^3 + 12L^2 s^2 - 60L + 120}{L^3 s^3 + 12L^2 s^2 + 60L + 120} \tag{4}$$

Here, the reduced order templates are given as:

$$P_{FOPTD}(s) = \frac{Ke^{-Ls}}{(\tau s + 1)} \tag{5}$$

$$P_{SOPTD}(s) = \frac{Ke^{-Ls}}{(\tau_{max} s + 1)(\tau_{min} s + 1)} \tag{6}$$

with the reduced order parameters $\{K, \tau, L\}$ denoting the dc-gain, time-constant (maximum or minimum) and time-delay respectively.

### 2.2. Model reduction of a test-bench of higher order processes:

In this paper, four set of test bench of higher order processes (7)-(10) have been studied as reported in Astrom & Hagglund [39]. Process $P_1$ represents a class of higher order processes with concurrent poles. Process $P_2$ represents a class of fourth order processes with increasing order of smallest time constants ($\alpha^3$). Process $P_3$ represents a class of third order processes with different values of the repeated dominant/non-dominant time constant ($T$). Process $P_4$ represents a class of non-minimum phase processes with increasing magnitude of the real right half plane zero.

$$P_1(s) = \frac{1}{(1+s)^n}, n \in \{3, 4, 5, 6, 7, 8, 10, 20\} \tag{7}$$

$$P_2(s) = \frac{1}{(1+s)(1+\alpha s)(1+\alpha^2 s)(1+\alpha^3 s)}, \tag{8}$$
$$\alpha \in \{0.1, 0.2, 0.3, 0.4, 0.5, 0.6, 0.7, 0.8, 0.9\}$$

$$P_3(s) = \frac{1}{(1+s)(1+sT)^2}, \tag{9}$$
$$T \in \{0.005, 0.01, 0.02, 0.05, 0.1, 0.2, 0.5, 2, 5, 10\}$$

$$P_4(s) = \frac{(1-\alpha s)}{(1+s)^3}, \tag{10}$$
$$\alpha \in \{0.1, 0.2, 0.3, 0.4, 0.5, 0.6, 0.7, 0.8, 0.9, 1.0, 1.1\}$$

The accuracies of the GA based optimization for model reduction using $H_2$-norm based and proposed Nyquist based technique approach has been compared in Fig. 1-4 and Table 1-2. Table 1-2 reports the sub-optimum reduced order FOPTD and SOPTD model parameters excluding the dc-gain ($K$) of the process which can be directly found out from the process model itself. It is clear that the proposed model reduction technique produces SOPTD models with high degree of accuracy in the Nyquist plane. Also, FOPTD models for the test-bench of higher order processes are less accurate than the SOPTD models with modeling objectives (1) and (3) respectively. In each case, accuracy of the FOPTD/SOPTD models with the proposed Nyquist based reduced order modeling technique is better than the $H_2$-norm based technique. It is also evident from Table 1-2 that the estimated delays in the GA based model reduction process is always lesser for the SOPTD models than the FOPTD models which gives much accurate compressed models.

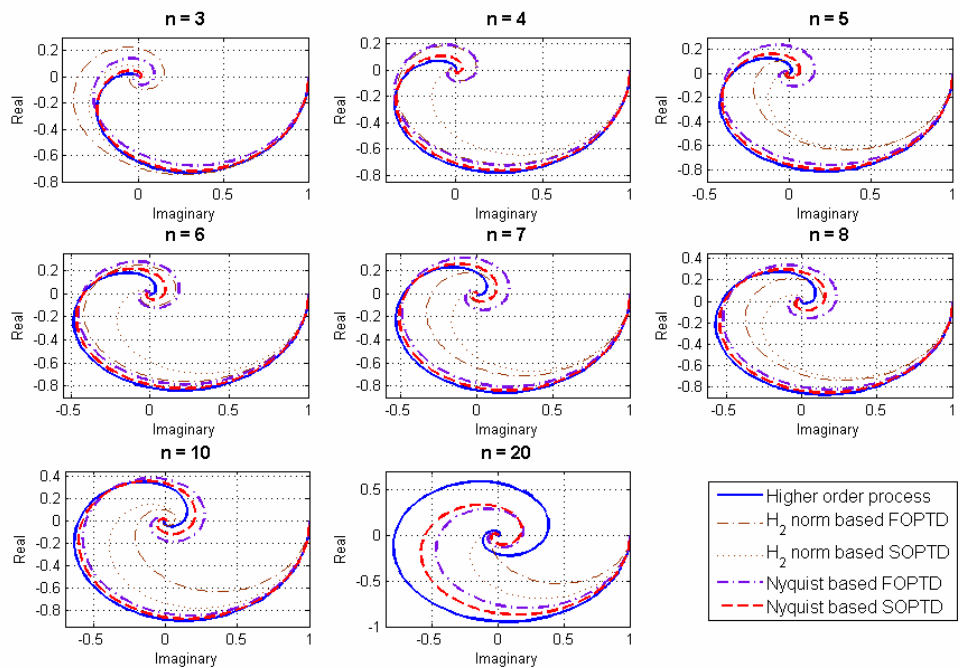

Fig. 1. Accuracies of reduced parameter models of $P_1$ in the Nyquist plane.

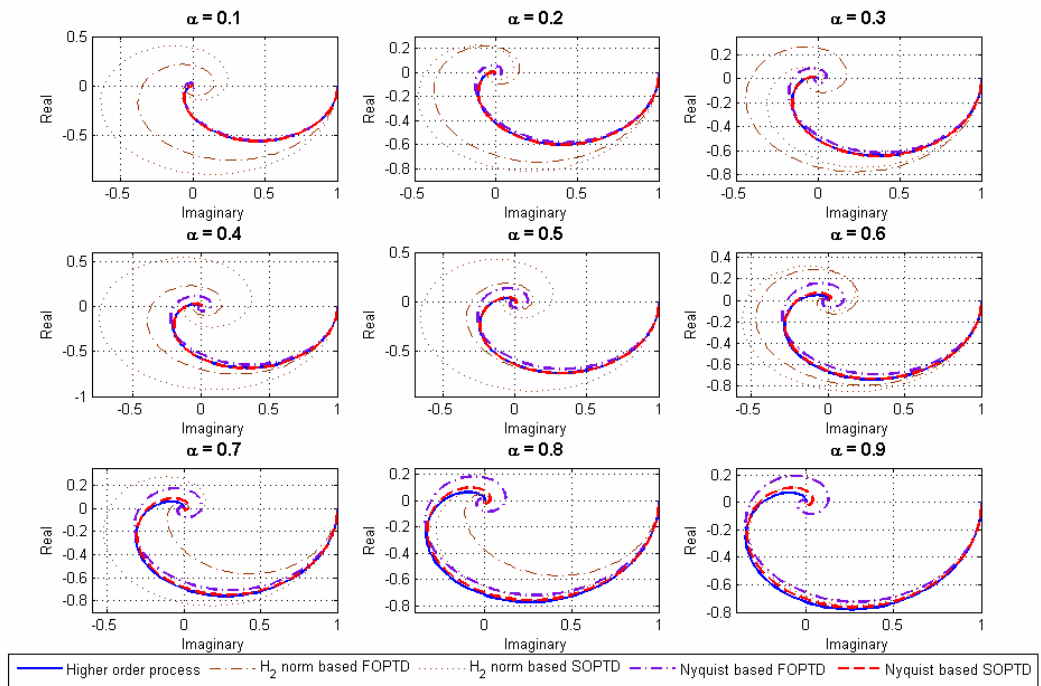

Fig. 2. Accuracies of reduced parameter models of $P_2$ in the Nyquist plane.

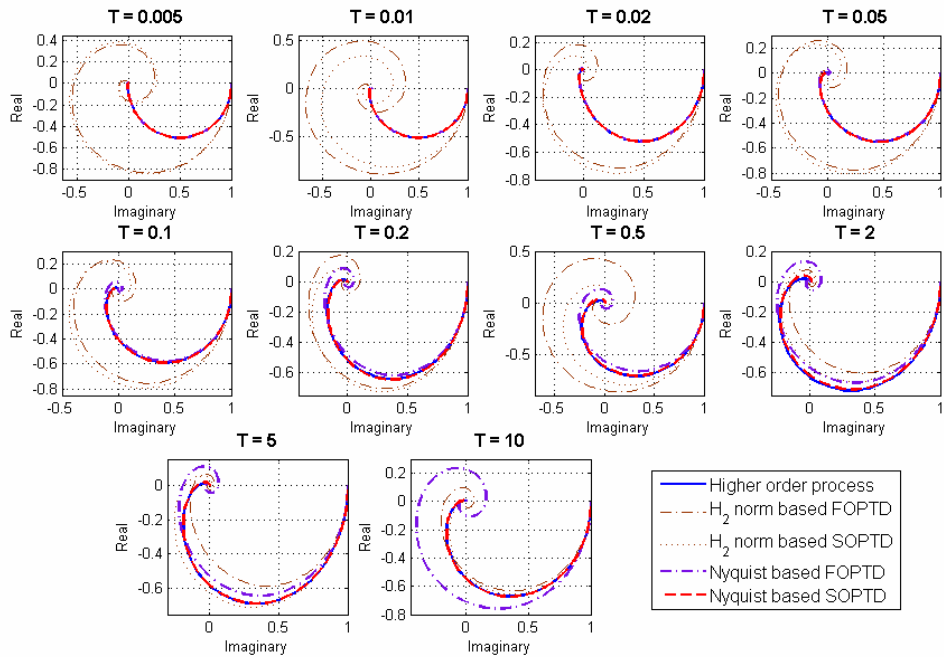

Fig. 3. Accuracies of reduced parameter models of $P_3$ in the Nyquist plane.

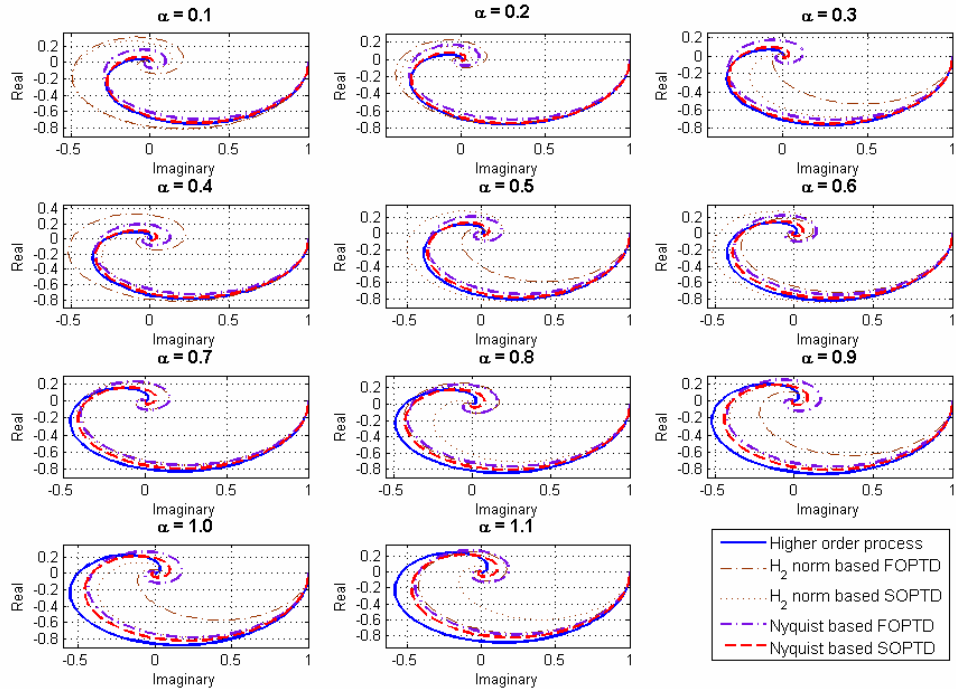

Fig. 4. Accuracies of reduced parameter models of $P_4$ in the Nyquist plane.

Table 1
$H_2$ norm based (1) reduced order FOPTD and SOPTD model parameters for the test-bench of higher-order processes

| Class of Processes | Varying Process Parameter | FOPTD Parameters | | | SOPTD Parameters | | | |
|---|---|---|---|---|---|---|---|---|
| | | $J_{min}$ | $\tau$ | L | $J_{min}$ | $\tau_{max}$ | $\tau_{min}$ | L |
| $P_1$ | n=3 | $7.27\times10^{-7}$ | 2.666658 | 1.99839 | $1.48\times10^{-7}$ | 1.832251 | 0.834414 | 0.734442 |
| | n=4 | $1.41\times10^{-7}$ | 3.199998 | 1.927009 | $1\times10^{-7}$ | 2.071363 | 1.128635 | 0.076068 |
| | n=5 | $1.66\times10^{-7}$ | 3.65714 | 1.217312 | $3.47\times10^{-9}$ | 2.661426 | 0.995717 | 0.101026 |
| | n=6 | $1.66\times10^{-7}$ | 4.063488 | 3.383379 | $2.13\times10^{-8}$ | 2.319241 | 1.74425 | 0.485802 |
| | n=7 | $1.15\times10^{-7}$ | 4.432903 | 2.629295 | $6.7\times10^{-8}$ | 2.568146 | 1.864757 | 0.273254 |
| | n=8 | $1.94\times10^{-7}$ | 4.773887 | 3.402165 | $3.65\times10^{-8}$ | 2.755938 | 2.017955 | 0.958559 |
| | n=10 | $5.02\times10^{-9}$ | 5.391691 | 1.773249 | $1.88\times10^{-9}$ | 3.130401 | 2.26129 | 2.392998 |
| | n=20 | $4.87\times10^{-8}$ | 7.776934 | 0.504645 | $9.44\times10^{-9}$ | 4.379741 | 3.397197 | 0.900555 |
| $P_2$ | α=0.1 | $3.32\times10^{-6}$ | 1.10099 | 0.802101 | $5.91\times10^{-7}$ | 0.631882 | 0.469117 | 1.273154 |
| | α=0.2 | $4.59\times10^{-7}$ | 1.208049 | 0.88278 | $9.76\times10^{-10}$ | 0.740946 | 0.467105 | 0.772794 |
| | α=0.3 | $4.1\times10^{-7}$ | 1.32751 | 1.199304 | $2.74\times10^{-7}$ | 0.875902 | 0.451609 | 0.394196 |
| | α=0.4 | $2.83\times10^{-7}$ | 1.466484 | 1.122393 | $1.17\times10^{-7}$ | 0.780577 | 0.685912 | 2.506013 |
| | α=0.5 | $3.98\times10^{-7}$ | 1.633067 | 0.999495 | $5.16\times10^{-8}$ | 0.955205 | 0.67786 | 2.041591 |
| | α=0.6 | $1.96\times10^{-7}$ | 1.836235 | 1.846691 | $7.04\times10^{-8}$ | 1.14874 | 0.687497 | 1.623046 |
| | α=0.7 | $7.34\times10^{-6}$ | 2.085431 | 0.305921 | $2.26\times10^{-7}$ | 1.1705 | 0.914927 | 1.5774 |
| | α=0.8 | $7.83\times10^{-6}$ | 2.389974 | 0.362211 | $9.57\times10^{-7}$ | 1.364212 | 1.02576 | 0.543319 |
| | α=0.9 | $1.18\times10^{-6}$ | 2.758773 | 1.723153 | $3.65\times10^{-7}$ | 1.817696 | 0.941083 | 0.886626 |
| $P_3$ | T=0.005 | $5.3\times10^{-6}$ | 1.007491 | 1.344538 | $2.4\times10^{-7}$ | 1.002721 | 0.004784 | 1.45683 |
| | T=0.01 | $2.64\times10^{-6}$ | 1.015032 | 2.070744 | $5.06\times10^{-7}$ | 0.968428 | 0.046596 | 1.109412 |
| | T=0.02 | $5.23\times10^{-7}$ | 1.030101 | 0.615312 | $1.05\times10^{-7}$ | 0.60851 | 0.421585 | 0.324902 |
| | T=0.05 | $3.96\times10^{-7}$ | 1.075609 | 0.953032 | $1.23\times10^{-7}$ | 0.777072 | 0.298542 | 0.682811 |
| | T=0.1 | $1.49\times10^{-6}$ | 1.152386 | 0.891076 | $8.9\times10^{-7}$ | 0.854932 | 0.297452 | 0.677295 |
| | T=0.2 | $2.03\times10^{-6}$ | 1.309099 | 0.727704 | $3.06\times10^{-7}$ | 0.893036 | 0.416054 | 0.344714 |
| | T=0.5 | $1.78\times10^{-7}$ | 1.800001 | 3.035848 | $4.91\times10^{-8}$ | 0.908001 | 0.891998 | 0.868299 |
| | T=2 | $2.98\times10^{-7}$ | 4.499992 | 1.056514 | $5.59\times10^{-8}$ | 3.892845 | 0.607153 | 1.031554 |
| | T=5 | 0.003128 | 9.999951 | 1.917817 | $1.87\times10^{-7}$ | 6.844189 | 3.441543 | 2.020844 |
| | T=10 | 0.066148 | 9.999989 | 3.115555 | 0.000757 | 9.996506 | 9.977595 | 0.365331 |
| $P_4$ | α=0.1 | $1.4\times10^{-7}$ | 2.657806 | 2.906337 | $4.32\times10^{-8}$ | 1.946295 | 0.711512 | 1.917234 |
| | α=0.2 | $6.08\times10^{-7}$ | 2.631586 | 1.986581 | $2.11\times10^{-7}$ | 1.900221 | 0.731355 | 1.09385 |
| | α=0.3 | $1.1\times10^{-6}$ | 2.588984 | 0.199346 | $9.65\times10^{-8}$ | 1.457495 | 1.1315 | 0.261696 |
| | α=0.4 | $4.04\times10^{-7}$ | 2.53165 | 2.989743 | $1.36\times10^{-7}$ | 1.622261 | 0.909386 | 0.72432 |
| | α=0.5 | $3.47\times10^{-7}$ | 2.461542 | 0.456537 | $1.03\times10^{-7}$ | 1.727013 | 0.734524 | 1.850298 |
| | α=0.6 | $1.13\times10^{-7}$ | 2.380954 | 1.427578 | $2.65\times10^{-8}$ | 1.433842 | 0.94711 | 1.776139 |
| | α=0.7 | $5.14\times10^{-7}$ | 2.292269 | 1.701698 | $1.47\times10^{-7}$ | 2.107201 | 0.185061 | 1.586829 |
| | α=0.8 | $5.91\times10^{-8}$ | 2.197803 | 1.935674 | $1.53\times10^{-8}$ | 1.320622 | 0.87718 | 0.38307 |
| | α=0.9 | $3.49\times10^{-7}$ | 2.099735 | 0.711619 | $1.25\times10^{-7}$ | 1.436491 | 0.663248 | 1.046835 |
| | α=1.0 | $1.43\times10^{-7}$ | 2.000001 | 0.303182 | $1.69\times10^{-8}$ | 1.000491 | 0.999509 | 0.774214 |
| | α=1.1 | $8.77\times10^{-7}$ | 1.900231 | 1.63462 | $4.38\times10^{-7}$ | 1.898203 | 0.002038 | 0.421641 |

Table 2
Nyquist based (3) reduced order FOPTD and SOPTD model parameters for the test-bench of higher-order processes

| Class of Processes | Varying Process Parameter | FOPTD Parameters | | | SOPTD Parameters | | | |
|---|---|---|---|---|---|---|---|---|
| | | $J_{min}$ | $\tau$ | L | $J_{min}$ | $\tau_{max}$ | $\tau_{min}$ | L |
| $P_1$ | n=3 | 1.317671 | 2.321831 | 1.035336 | 0.35763 | 1.335035 | 1.296596 | 0.458524 |
| | n=4 | 1.458225 | 2.746493 | 1.73781 | 0.534457 | 1.586542 | 1.548473 | 1.03317 |
| | n=5 | 1.546008 | 3.109802 | 2.485838 | 0.643986 | 1.797635 | 1.770904 | 1.666146 |
| | n=6 | 1.610808 | 3.432621 | 3.26406 | 0.720594 | 1.989875 | 1.959647 | 2.344943 |
| | n=7 | 1.665762 | 3.727855 | 4.063357 | 0.779376 | 2.163055 | 2.14323 | 3.051016 |
| | n=8 | 1.71735 | 4.003052 | 4.878174 | 0.82832 | 2.310304 | 2.310215 | 3.782639 |
| | n=10 | 1.821824 | 4.514103 | 6.54117 | 0.91604 | 2.661457 | 2.549809 | 5.293009 |
| | n=20 | 4.480042 | 9.999542 | 9.99943 | 2.504335 | 5.451683 | 5.397813 | 9.999728 |
| $P_2$ | α=0.1 | 0.344204 | 1.038803 | 0.091833 | 0.004308 | 0.999772 | 0.100915 | 0.010279 |
| | α=0.2 | 0.594509 | 1.10725 | 0.192456 | 0.028107 | 0.992451 | 0.214076 | 0.038794 |
| | α=0.3 | 0.792144 | 1.193443 | 0.311536 | 0.060572 | 0.979505 | 0.341498 | 0.092874 |
| | α=0.4 | 0.960828 | 1.299211 | 0.453122 | 0.107937 | 0.943464 | 0.51063 | 0.167586 |
| | α=0.5 | 1.109156 | 1.430175 | 0.618736 | 0.173435 | 0.833884 | 0.778235 | 0.270018 |
| | α=0.6 | 1.236298 | 1.594346 | 0.807116 | 0.292888 | 0.919789 | 0.886179 | 0.409777 |
| | α=0.7 | 1.337076 | 1.800325 | 1.015772 | 0.400586 | 1.026115 | 1.021073 | 0.559864 |
| | α=0.8 | 1.407224 | 2.056009 | 1.242017 | 0.480812 | 1.233382 | 1.10547 | 0.720248 |
| | α=0.9 | 1.446406 | 2.369119 | 1.48327 | 0.521566 | 1.371358 | 1.331686 | 0.879882 |
| $P_3$ | T=0.005 | 0.029338 | 1.000578 | 0.009644 | 0.003451 | 1.000027 | 0.007301 | 0.00276 |
| | T=0.01 | 0.058209 | 1.001793 | 0.019079 | 0.006693 | 0.999721 | 0.014931 | 0.005228 |
| | T=0.02 | 0.114515 | 1.00517 | 0.037582 | 0.013254 | 0.999557 | 0.030272 | 0.010203 |
| | T=0.05 | 0.271088 | 1.021791 | 0.090546 | 0.031173 | 0.997605 | 0.075538 | 0.026398 |
| | T=0.1 | 0.491118 | 1.063062 | 0.171636 | 0.05823 | 0.989257 | 0.157307 | 0.050227 |
| | T=0.2 | 0.805051 | 1.172148 | 0.313077 | 0.100513 | 0.963887 | 0.337572 | 0.09348 |
| | T=0.5 | 1.208464 | 1.574149 | 0.639661 | 0.243507 | 0.911085 | 0.868222 | 0.253221 |
| | T=2 | 1.242701 | 3.932746 | 1.628066 | 0.274858 | 2.285902 | 2.162089 | 0.662506 |
| | T=5 | 1.097098 | 9.083402 | 3.023718 | 0.105979 | 5.271248 | 4.954549 | 0.85439 |
| | T=10 | 2.907921 | 9.999917 | 7.937636 | 0.048469 | 9.999702 | 9.998882 | 0.98878 |
| $P_4$ | α=0.1 | 1.311728 | 2.317266 | 1.135967 | 0.350007 | 1.321307 | 1.304839 | 0.562264 |
| | α=0.2 | 1.29935 | 2.305755 | 1.235753 | 0.334032 | 1.317905 | 1.293675 | 0.66746 |
| | α=0.3 | 1.286855 | 2.289458 | 1.33284 | 0.332085 | 1.393695 | 1.197571 | 0.773718 |
| | α=0.4 | 1.282005 | 2.269393 | 1.426222 | 0.351824 | 1.334063 | 1.234247 | 0.873208 |
| | α=0.5 | 1.293081 | 2.246754 | 1.515202 | 0.423653 | 1.298311 | 1.242496 | 0.968798 |
| | α=0.6 | 1.327608 | 2.222476 | 1.599412 | 0.542731 | 1.25362 | 1.252805 | 1.064005 |
| | α=0.7 | 1.390988 | 2.196868 | 1.678904 | 0.698068 | 1.241163 | 1.240979 | 1.150465 |
| | α=0.8 | 1.485649 | 2.170445 | 1.753729 | 0.881815 | 1.293128 | 1.161037 | 1.234179 |
| | α=0.9 | 1.611152 | 2.143507 | 1.824189 | 1.085803 | 1.28306 | 1.138877 | 1.308246 |
| | α=1.0 | 1.765058 | 2.116299 | 1.89035 | 1.307159 | 1.298524 | 1.09749 | 1.387555 |
| | α=1.1 | 1.943976 | 2.088913 | 1.952693 | 1.542905 | 1.312971 | 1.053957 | 1.459166 |

It has been shown in Fig. 1-4 that the Nyquist based model reduction produces better quality of compresses models than with the $H_2$ norm based one. This is evident from Fig. 1-4 as the Nyquist based SOPTD models closely follows the original higher order process in each case than the other three cases. To further justify the point, a comparison of the achievable accuracies of the $H_2$ norm based and proposed Nyquist based model reduction technique has been shown on the basis of objective function (3) and shown in semi-log/log-log scale in Fig 5. It is clear from Fig. 5 that for each case the proposed Nyquist based SOPTD models yields more accurate models in frequency domain over that with the $H_2$ norm based methods.

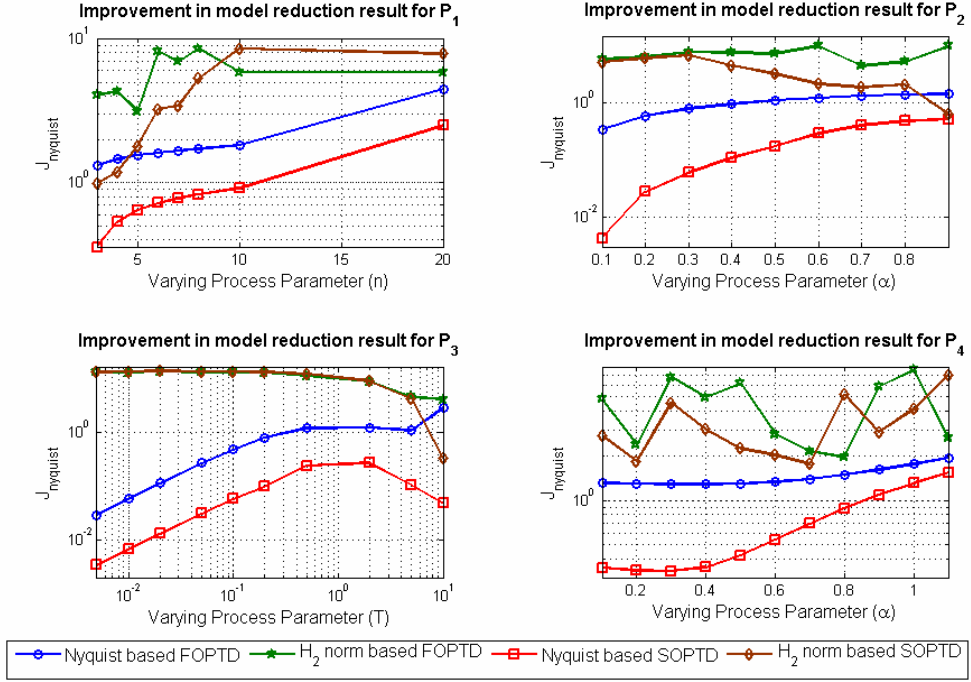

Fig. 5. Model reduction errors for the proposed Nyquist based and $H_2$-norm based FOPTD/SOPTD models.

## 3. Generation of time domain optimal controller tuning rule:
### 3.1. Controller structure and objective function for tuning:

In this paper, the performance of two classes of controllers has been studied to control few higher order processes given by (7)-(10). The chosen controllers are conventional PID type which is widely used in process control industries and its analogous fractional order $PI^\lambda D^\mu$, proposed by Podlubny [21] which is gaining increased interest amongst the research community [22]-[26]. The $PI^\lambda D^\mu$ controller has been considered to have a parallel structure (11) similar to the conventional PID controller [1].

$$C_{FOPID}(s) = K_p + \frac{K_i}{s^\lambda} + K_d s^\mu \qquad (11)$$

Clearly, the $PI^\lambda D^\mu$ controller (11) is a generalization of the classical PID controller with two extra tuning knob i.e. the integro-differential orders $\{\lambda, \mu\}$. The conventional PID controller can be designed with the same technique by putting $\{\lambda, \mu\} = 1$. The PID and $PI^\lambda D^\mu$ controllers are now tuned with a constrained Genetic Algorithm, since its unconstrained version may produce large controller gains and increase the cost of hardware implementation. The goal of the constrained optimization is to minimize a weighted sum of a suitable error index and the controller effort (12) similar to that in Pan *et al.* [11]-[12]:

$$J = \int_0^\infty \left[ w_1 \cdot t \cdot |e(t)| + w_2 \cdot u^2(t) \right]$$  (12)

Here, the first term corresponds to the ITAE which minimizes the overshoot and settling time, whereas the second term denotes the Integral of Squared Controller Output (ISCO). The two weights $\{w_1, w_2\}$ balances the impact of control loop error (oscillation and/or sluggishness) and control signal (larger actuator size and chance of integral wind-up) and both have been chosen to be unity in the present simulation study indicating same penalty for large magnitude of ITAE and ISCO.

### *3.2. Application of genetic algorithm for optimal controller tuning:*

Genetic Algorithm is a computational stochastic method for optimization based on the natural Darwinian evolution. In GA each solution vector (chromosome) is represented by real valued bit strings which are essentially an encoded form of the solution variables. These chromosomes evolve over successive generations through evolutionary operations like reproduction, crossover and mutation. Each set of solution vector in the mating pool is assigned a relative fitness value based on the evaluation of an objective function. A scaling function is converts the raw fitness scores in a form that is suitable for the selection function. Rank fitness scaling is used which scales the raw scores on the basis of its position in the sorted score list. This removes the effect of the spread of the raw scores. The fitter individuals have a greater probability of passing on to the next generation. Newer individuals are created on probabilistic decisions from parent genes by the process of crossover. A scattered crossover function is used which creates a random binary vector and selects the genes where the vector has a value of 1 from the first parent, and the genes where the vector has a value of 0 from the second parent, and combines the genes to form the child. Mutation is applied at randomly selected positions of the parent gene to produce newer individuals. For mutation the Gaussian function is used which adds a random number to each vector entry of an individual. This random number is taken from a Gaussian distribution centered around zero. With these operators newer individuals are produced and the solution is iteratively refined until the objective function is minimized below a certain tolerance level or the maximum number of iterations are exceeded.

Another parameter called the elite count is also used in the GA. This represents the number of fittest individuals in the present generation which will definitely be copied over to the next generation. Usually this number is small, as otherwise the initially obtained fitter individuals would dominate and would lead to premature convergence of the algorithm. The number of individuals other than the elite, in the present generation, that evolve through crossover and the number that evolve through mutation are pre-specified by the crossover fraction and the mutation fraction respectively. In this case the mutation fraction is chosen to be 0.2 and the crossover fraction as 0.8. The GA population is chosen to be 20 and the elite count as 2. The selection function chooses the vectors which act as parents of the next generation based on the inputs from the fitness scaling function. Here a stochastic uniform function is used. These values have been adopted since they have proved effective in a wide variety of optimization problems [11]-[14]. Also for controller tuning problem, the objective function evaluation is computationally intensive and hence a rigorous parametric variation of the GA is beyond the scope of the present work.

The variables that constitute the search space for the PID and the fractional order $PI^\lambda D^\mu$ controller are $\{K_p, K_i, K_d\}$ and $\{K_p, K_i, K_d, \lambda, \mu\}$ respectively. The intervals of the search space for these variables are $\{K_p, K_i, K_d\} \in [0,100]$ and $\{\lambda, \mu\} \in [0,2]$. The present problem searches for the optimal controller parameters while minimizing the time domain integral performance index (12). The corresponding optimal controller parameters are reported in Table 3 and 4 respectively along with the minima of the control objective (12). Also, within the GA based optimization framework, the objective function (12) has been evaluated with a finite time horizon of 100 seconds.

### *3.3. Genetic programming based analytical tuning rule extraction for PID/$PI^\lambda D^\mu$ controllers:*

The GA based $PI^\lambda D^\mu$ controller parameters $\{K_p, K_i, K_d, \lambda, \mu\}$ and PID controller parameters $\{K_p, K_i, K_d\}$ are now used as the test data-set to develop an optimal tuning rule for the respective controllers with minimum value of the objective function gievn by (12). In this paper, Genetic Programming is used to optimally map the enhanced sub-optimal reduced order SOPTD parameters representing the higher order systems (Table 2) and the optimal PID/$PI^\lambda D^\mu$ controller parameters (Table 3-4) for extracting optimal tuning rules based on the control objective (12), also in an optimum fashion. For tuning rule development several measures of standard SOPTD templates like time-delay ($L$), maximum-minimum time-constant ratio ($\tau_{max}/\tau_{min}$), time-delay to time constant ratio ($L/\tau_{min}$ and $L/\tau_{max}$) etc. have been used to map the GA based sub-optimal reduced order SOPTD parameters with GA based PID/FOPID parameters. O' Dwyer in [1] has reported least square based empirical rule extraction approach to fit a chosen structure of the tuning-rule. The idea has been improved in this paper with a GP based approach with optimal choice of the structure to fit the GA based optimized reduced model and controller data in the rule and with additional choice of the complexity, representing the tuning formula. In the present study, single-gene and multi-gene approaches of GP both have been used for optimum PID/FOPID tuning rule development. The single gene GP rules represent each $PI^\lambda D^\mu$ controller parameters (gains and orders) as nonlinear functions of reduced process parameters (dc-gain, delay and time constants) and the multi-gene rules represent linear combinations of the nonlinear functions of the respective SOPTD parameters. Fixed structure based FOPID tuning rule has been attempted in [28]-[33]. Whereas this paper proposes a new approach of process and controller data based automatic rule generation via GP. It is worth mentioning that early researches shows that GP is capable of producing human competitive PID like controller topology along with its parameters and successfully applied in process control applications like [40]-[50].

Table 3
Optimal PID controller tuning results for the test-bench of higher order processes

| Processes | Varying Process Parameter | PID Controller Parameters | | | |
|---|---|---|---|---|---|
| | | $J_{min}$ | $K_p$ | $K_i$ | $K_d$ |
| $P_1$ | n=3 | 104.9453 | 1.182448 | 0.413749 | 0.782454 |
| | n=4 | 108.3718 | 1.637145 | 0.354506 | 1.741753 |
| | n=5 | 112.6885 | 1.100964 | 0.237852 | 1.402792 |
| | n=6 | 119.4189 | 0.884417 | 0.183229 | 1.248403 |
| | n=7 | 128.0275 | 0.760101 | 0.147847 | 1.185624 |
| | n=8 | 138.054 | 0.716967 | 0.125966 | 1.304152 |
| | n=10 | 164.4938 | 0.701403 | 0.099816 | 1.783894 |
| | n=20 | 389.3276 | 0.494927 | 0.044386 | 1.921364 |
| $P_2$ | α=0.1 | 101.0431 | 0.796559 | 0.878393 | 0.010446 |
| | α=0.2 | 101.0577 | 1.215855 | 0.869939 | 0.500778 |
| | α=0.3 | 101.4173 | 0.833683 | 0.777554 | 0.060493 |
| | α=0.4 | 101.7203 | 0.840911 | 0.687642 | 0.108355 |
| | α=0.5 | 102.179 | 0.935624 | 0.62135 | 0.231097 |
| | α=0.6 | 102.8128 | 0.973036 | 0.554198 | 0.311916 |
| | α=0.7 | 103.5996 | 1.000883 | 0.470891 | 0.437664 |
| | α=0.8 | 104.8522 | 0.860593 | 0.374364 | 0.429814 |
| | α=0.9 | 106.2945 | 0.969693 | 0.33737 | 0.635315 |
| $P_3$ | T=0.005 | 100.9342 | 0.78585 | 0.8808 | 0.060011 |
| | T=0.01 | 100.8663 | 1.142077 | 0.988279 | 0.341946 |
| | T=0.02 | 100.9604 | 0.978586 | 1.001575 | 0.096083 |
| | T=0.05 | 101.0173 | 0.799997 | 0.884878 | 0.010097 |
| | T=0.1 | 101.1476 | 0.755547 | 0.792823 | 0.042604 |
| | T=0.2 | 101.4033 | 0.854037 | 0.776625 | 0.085347 |
| | T=0.5 | 102.4682 | 0.974437 | 0.602782 | 0.234556 |
| | T=2 | 111.3098 | 1.511128 | 0.297816 | 1.393981 |
| | T=5 | 142.1802 | 1.993027 | 0.170803 | 3.388806 |
| | T=10 | 241.7192 | 1.823249 | 0.08681 | 4.057907 |
| $P_4$ | α=0.1 | 105.2964 | 1.039593 | 0.377487 | 0.675297 |
| | α=0.2 | 105.3511 | 1.22371 | 0.401408 | 0.867481 |
| | α=0.3 | 105.6178 | 1.121824 | 0.378009 | 0.785138 |
| | α=0.4 | 105.8309 | 1.264824 | 0.388351 | 0.970807 |
| | α=0.5 | 106.0636 | 1.093624 | 0.355008 | 0.818021 |
| | α=0.6 | 106.3366 | 1.023166 | 0.337257 | 0.764305 |
| | α=0.7 | 106.7822 | 1.27256 | 0.361916 | 1.084575 |
| | α=0.8 | 106.8299 | 1.008027 | 0.321601 | 0.792297 |
| | α=0.9 | 107.1832 | 0.893576 | 0.298821 | 0.681007 |
| | α=1.0 | 107.4011 | 0.91616 | 0.296311 | 0.728001 |
| | α=1.1 | 107.9277 | 0.788123 | 0.274588 | 0.587885 |

Table 4
Optimal $PI^\lambda D^\mu$ controller tuning results for the test-bench of higher order processes

| Processes | Varying Process Parameter | $PI^\lambda D^\mu$ Controller Parameters | | | | | |
|---|---|---|---|---|---|---|---|
| | | $J_{min}$ | $K_p$ | $K_i$ | $K_d$ | $\lambda$ | $\mu$ |
| $P_1$ | n=3 | 105.7456 | 0.567381 | 0.397193 | 0.336985 | 0.997252 | 0.238964 |
| | n=4 | 109.9111 | 0.630373 | 0.2941 | 0.364797 | 0.996471 | 0.537579 |
| | n=5 | 115.5586 | 0.593071 | 0.215924 | 0.327656 | 0.998192 | 0.658275 |
| | n=6 | 123.8322 | 0.479687 | 0.184532 | 0.437726 | 0.994454 | 0.465764 |
| | n=7 | 133.0418 | 0.542002 | 0.175413 | 0.775213 | 0.9946 | 0.632662 |
| | n=8 | 141.6016 | 0.48947 | 0.129047 | 0.488187 | 0.995167 | 0.66905 |
| | n=10 | 165.6369 | 0.486645 | 0.109355 | 0.754894 | 0.991345 | 0.707808 |
| | n=20 | 372.7587 | 0.248849 | 0.077716 | 1.667382 | 0.902758 | 0.654193 |
| $P_2$ | α=0.1 | 101.223 | 1.313521 | 1.620008 | 0.291235 | 0.985475 | 0.274916 |
| | α=0.2 | 101.2217 | 0.556556 | 1.25093 | 0.710321 | 0.988136 | 0.111315 |
| | α=0.3 | 101.6836 | 0.915874 | 0.973408 | 0.161563 | 0.999982 | 0.212506 |
| | α=0.4 | 102.0359 | 0.88805 | 0.863005 | 0.192425 | 0.999348 | 0.377857 |
| | α=0.5 | 102.6392 | 0.518037 | 0.720932 | 0.485413 | 0.998278 | 0.110046 |
| | α=0.6 | 103.8012 | 0.635 | 0.667918 | 0.508921 | 0.999076 | 0.344127 |
| | α=0.7 | 104.1065 | 0.66709 | 0.460539 | 0.225573 | 0.998766 | 0.514388 |
| | α=0.8 | 105.9693 | 0.765149 | 0.458229 | 0.418873 | 0.998452 | 0.573715 |
| | α=0.9 | 107.5504 | 0.664219 | 0.358983 | 0.401456 | 0.997591 | 0.5517 |
| $P_3$ | T=0.005 | 101.834 | 0.649273 | 3.555929 | 0.986111 | 0.966164 | 0.116951 |
| | T=0.01 | 100.925 | 0.91604 | 1.183733 | 0.257541 | 0.988467 | 0.317111 |
| | T=0.02 | 101.5364 | 0.825772 | 2.005661 | 0.988452 | 0.995185 | 0.052581 |
| | T=0.05 | 101.4871 | 0.395052 | 1.584049 | 1.079818 | 0.998711 | 0.002762 |
| | T=0.1 | 101.448 | 1.051311 | 1.168991 | 0.063764 | 0.997074 | 0.108792 |
| | T=0.2 | 101.4178 | 0.863862 | 0.856676 | 0.012141 | 0.999923 | 0.434871 |
| | T=0.5 | 103.0134 | 0.371651 | 0.683201 | 0.676411 | 0.996869 | 0.131265 |
| | T=2 | 113.9015 | 1.049736 | 0.319246 | 0.679637 | 0.996232 | 0.568066 |
| | T=5 | 144.3692 | 1.069802 | 0.192226 | 1.529542 | 0.97701 | 0.436781 |
| | T=10 | 214.2168 | 0.813306 | 0.142782 | 3.101702 | 0.928298 | 0.403031 |
| $P_4$ | α=0.1 | 107.2669 | 0.594065 | 0.532913 | 0.834404 | 0.99512 | 0.352709 |
| | α=0.2 | 107.1442 | 0.703495 | 0.472681 | 0.564386 | 0.997582 | 0.418784 |
| | α=0.3 | 106.5329 | 0.69113 | 0.352834 | 0.249862 | 0.998902 | 0.606224 |
| | α=0.4 | 108.652 | 0.412859 | 0.461503 | 0.884647 | 0.995174 | 0.336957 |
| | α=0.5 | 108.5548 | 0.595193 | 0.42152 | 0.670115 | 0.996705 | 0.453979 |
| | α=0.6 | 108.9315 | 0.691254 | 0.413996 | 0.567369 | 0.997532 | 0.541673 |
| | α=0.7 | 108.9609 | 0.708224 | 0.378196 | 0.528091 | 0.998144 | 0.615374 |
| | α=0.8 | 112.3442 | 0.066768 | 0.45611 | 1.203985 | 0.992998 | 0.296968 |
| | α=0.9 | 109.2519 | 0.65186 | 0.330515 | 0.423453 | 0.997734 | 0.643539 |
| | α=1.0 | 110.0062 | 0.520752 | 0.317104 | 0.466304 | 0.996656 | 0.504782 |
| | α=1.1 | 112.426 | 0.52888 | 0.367182 | 0.706627 | 0.997325 | 0.562432 |

Genetic programming [40] is a class of computational intelligence techniques which extends the notion of the conventional Genetic Algorithm, to evolve computer programs which can perform user defined tasks. It is an evolutionary algorithm and is based on the biological strategies of reproduction, crossover and mutation to evolve fitter solutions in the future generations. In the present paper, GP is used for symbolic regression to find out an analytic expression that maps the input variables of the process parameters to the output values of the controller parameters while minimizing the mean absolute error (MAE) of the predicted controller parameters (from the rule) and the specified well-tuned values. Thus instead of finding the coefficients of a particular structure as in the conventional regression in [1], [28]-[33]; GP searches in the infinite dimensional functional space to find an optimum structure along with the numerical coefficients, minimizing MAE of the controller parameters.

In GP each candidate solution is a function itself and is encoded in the form of a tree. Fig. 6 shows the schematic for crossover between the two parent genes. Since the whole node with its corresponding sub-nodes get replaced in this case, so the crossover procedure is more effective and can provide a wide variety of individuals. Care must be taken so that the crossover process does not produce an indeterminate function or ill conditioned expression (e.g. division by zero, logarithm of a negative number etc.) and such solutions must be eliminated [51]. Fig. 7 shows the mutation schematic where a randomly chosen node in the tree is replaced by another randomly generated sub-tree giving rise to a new individual. For the present study the population size is chosen to be 500. A tournament selection method is adopted and the tournament size is kept as 3. The maximum depth of each tree is assumed to be 7. The set of functions used for symbolic regression are $\left\{+,-,\times,\div,\sqrt[n]{\ },\sin,\cos,\tanh,\log,e^x,square\right\}$. The crossover probability has been taken as 0.85, mutation probability as 0.1 and direct reproduction as 0.05.

Also, multi-gene symbolic regression has been shown to be more accurate and computationally efficient than the standard GP approach in [51]. Unlike the traditional single-gene GP approach, the multi-gene symbolic regression is the weighted linear combination of the outputs from a number of GP trees and each of these trees represent an individual gene. For each model the linear coefficients are estimated from the training data using standard least square techniques. The depth of each tree can be specified to lower values so as to restrict the complexity of the expressions. In multi-gene regression, apart from the standard methods of mutation and crossover, a two point high level crossover is also possible. This allows exchange of whole genes between two different individuals. The standard crossover operator is thus known as sub-tree crossover in this case as only a randomly selected part of the tree participates in the crossover and not the whole tree itself. For the multi-gene symbolic regression additionally the following parameters are used. The high level crossover has been taken to be 0.2, the low level crossover as 0.8 and the sub-tree mutation as 0.9 as suggested in [51].

Fig. 8-9 shows the Pareto optimal front for the fitness values versus the number of terms of the expression found from GP where each dot represents a solution expression with different level of complexity and fitness value. The circles indicate the non-Pareto optimal solutions and the down-head triangles indicate the Pareto optimal front. The solution having the lowest fitness has been highlighted with a star while the corresponding controller parameter accuracies have been shown in Fig. 10-11. It is obvious that the increase in the number of nodes increases the complexity of the overall

expression, but gives a better fit, i.e. a lower value of fitness function. However for ease of computability a trade-off can be made between the fitness and complexity by intuitive judgment as reported in [34].

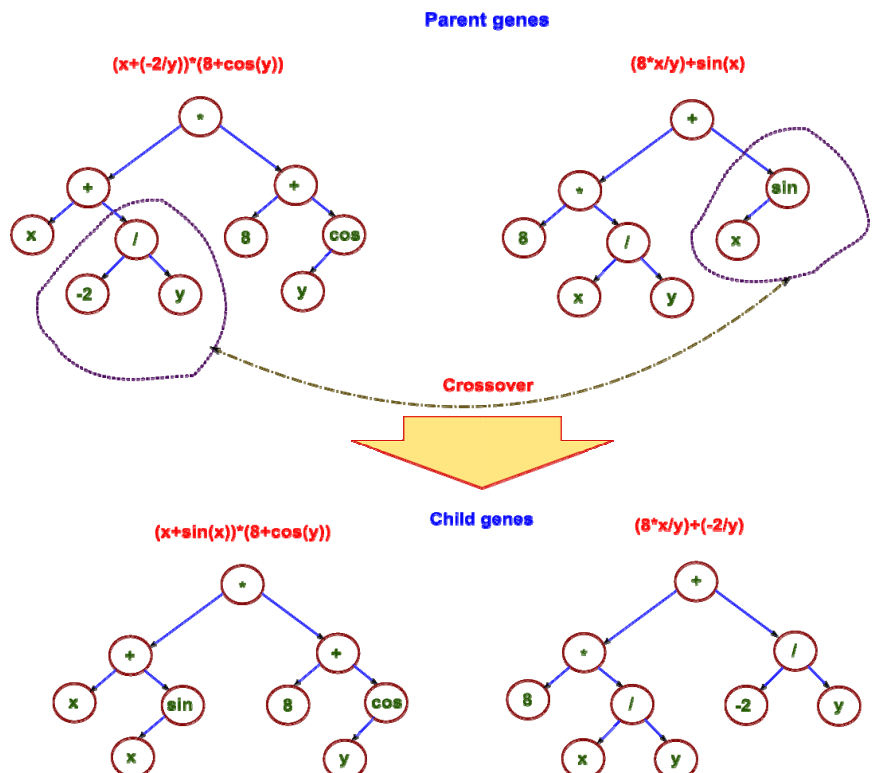

Fig. 6. Schematic of cross-over in genetic programming.

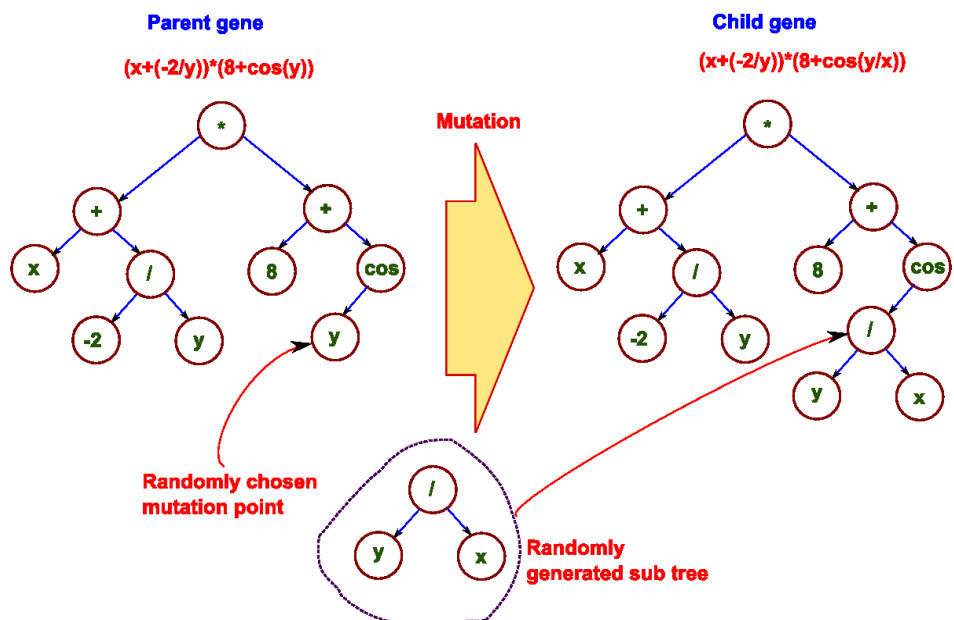

Fig. 7. Schematic of mutation in Genetic Programming.

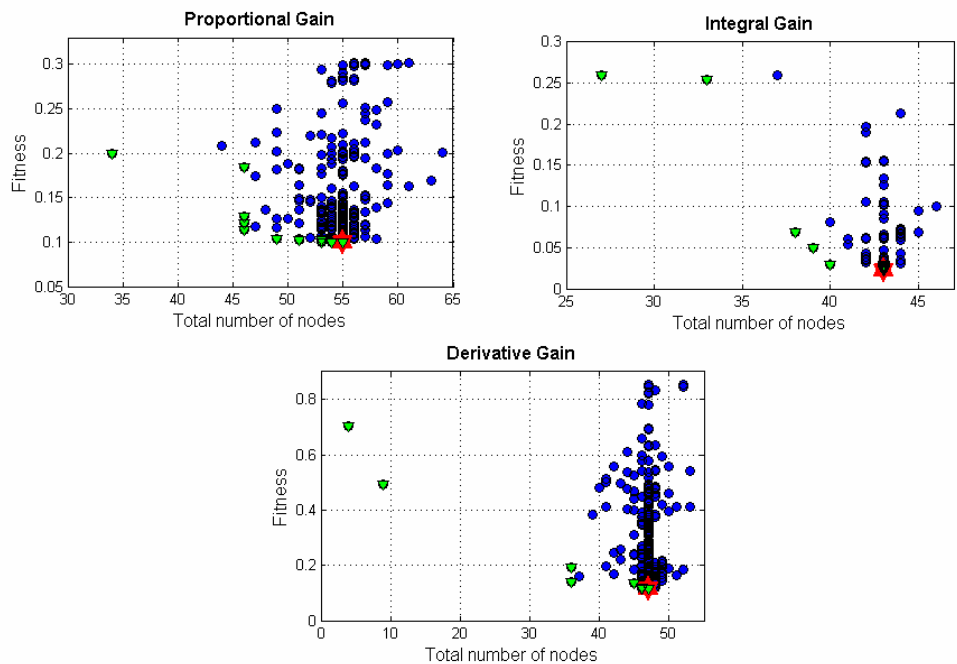

Fig. 8. Pareto optimal front showing fitness vs. complexity for the single gene PID controller tuning rule.

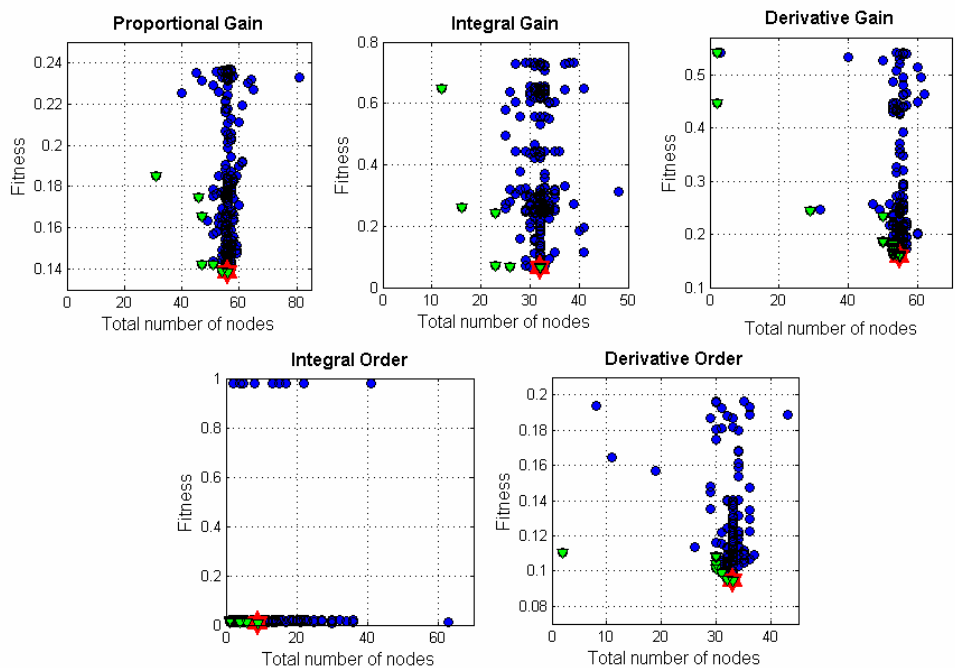

Fig. 9. Pareto optimal front showing fitness vs. complexity for the single gene $PI^\lambda D^\mu$ controller tuning rule.

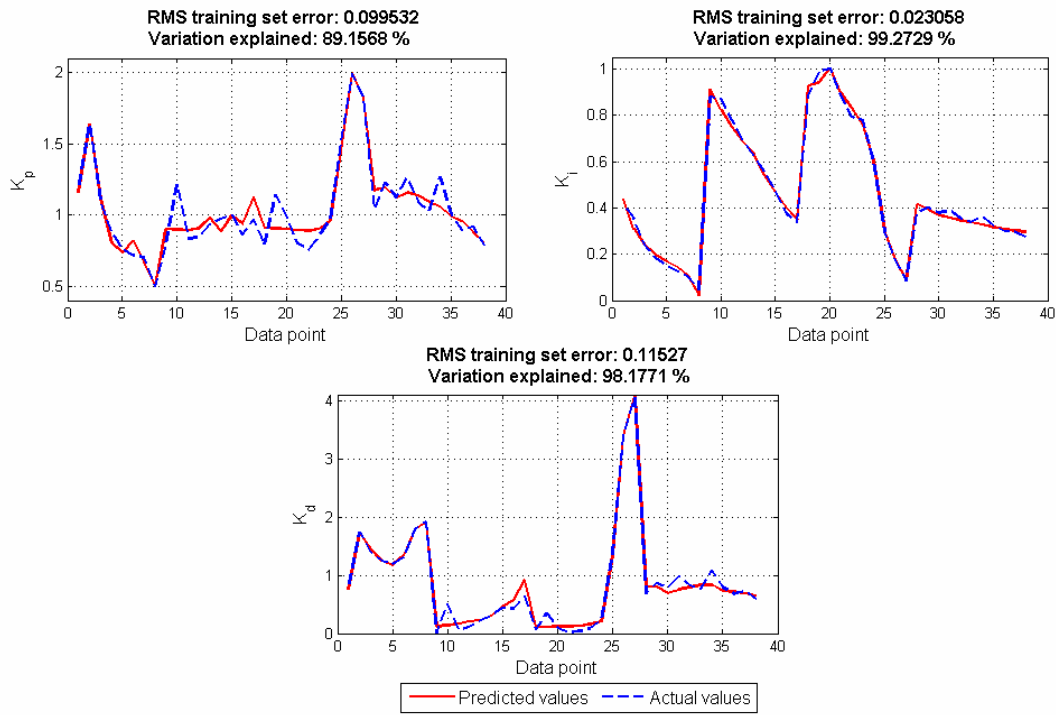

Fig. 10. Accuracies of the different PID controller parameters with the best found single gene tuning rule.

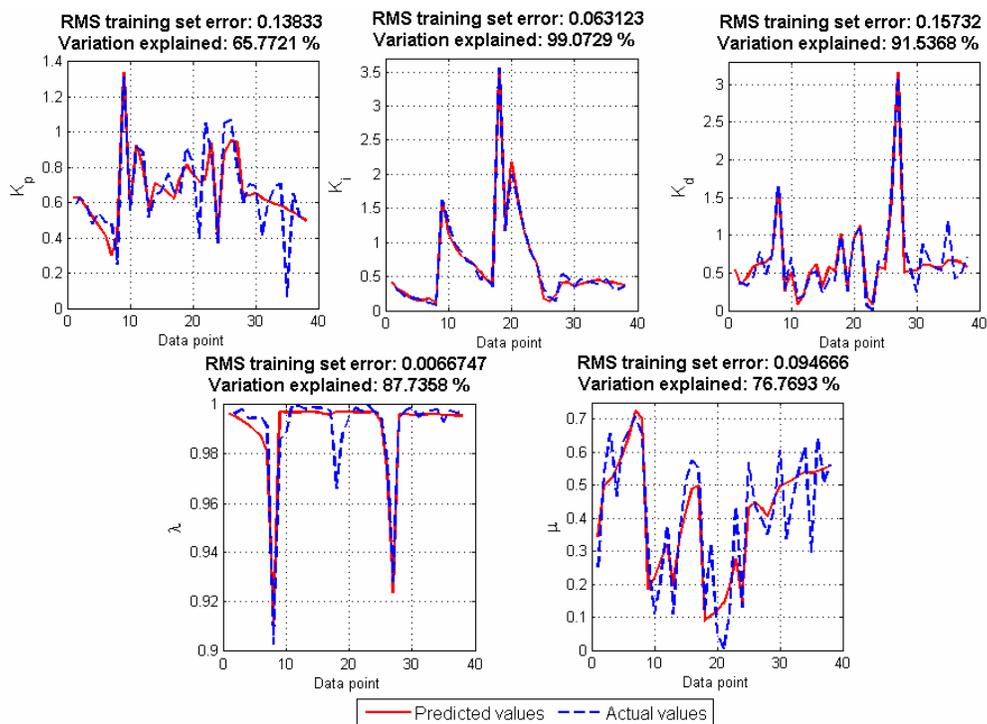

Fig. 11. Accuracies of the different $PI^\lambda D^\mu$ controller parameters with the best found single gene tuning rule.

The Pareto fronts for the single and multi-gene cases have been shown in Figs. 8-9 and 12-13 respectively. For single gene case, the total number of nodes of the expression for the best fit gene varies approximately between 40 to 55 for the PID parameters and between 30 to 60 for the FOPID parameters. The nodes for the best fit gene of the derivative and integral orders of the FOPID controller are comparatively less indicating that the variation of the orders are less than the gains. However for the multi-gene case, the number of nodes for the gains of the PID controller varies approximately between 80 to 150 which are much higher than the corresponding single gene cases. With multi-gene GP, the number of nodes for the expressions of the gains and orders of the FOPID controller varies between 100 to 190 and 40 to 70 respectively which also indicate a higher degree of complexity than the single gene expressions. Thus although the multi-gene models may be capable of explaining the nonlinear mapping between the SOPTD reduced order model parameters $\{K, \tau_{max}, \tau_{min}, L\}$ and PID/FOPID parameters $\{K_p, K_i, K_d, \lambda, \mu\}$ to a greater extent, the complexity of the rules make them a huge impediment towards actual implementation in real time automation.

The multi-gene symbolic regression approach show better fit of the controller parameters than their single gene counterpart as is evident from Figs. 14-15. However, the complexity of the expressions increases drastically. However, if accuracy is more important for some application and practical hardware issues can be surmounted then this approach can give better control system performance at the cost of increased computational complexity. Also, multi-gene GP rules are preferable as they more accurately maps the GA based tuning results of optimum controller parameters. Even the best found single-gene GP based results give less accurate but low complexity rules. Applications with strictly requirement of low complexity rules are referred to [34].

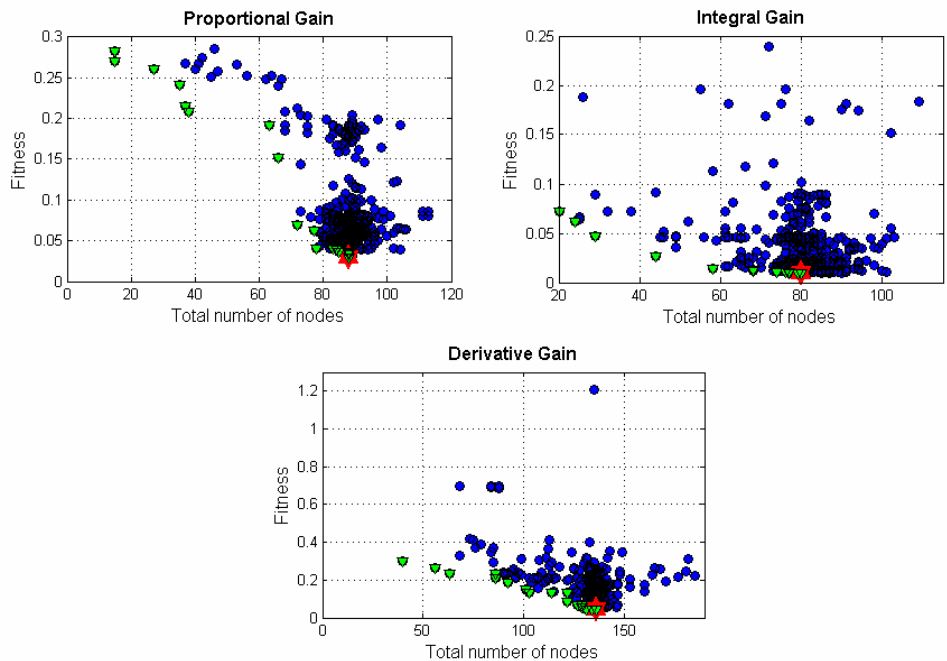

Fig. 12. Pareto optimal front showing fitness vs. complexity for the multi-gene PID controller tuning rule.

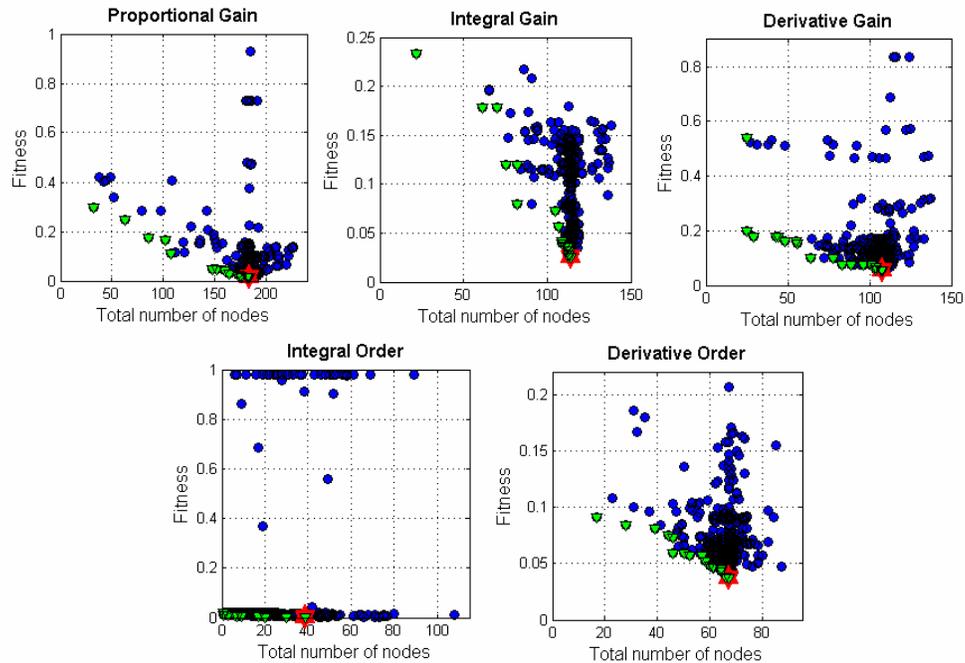

Fig. 13. Pareto optimal front showing fitness vs. complexity for the multi-gene $PI^{\lambda}D^{\mu}$ controller tuning rule.

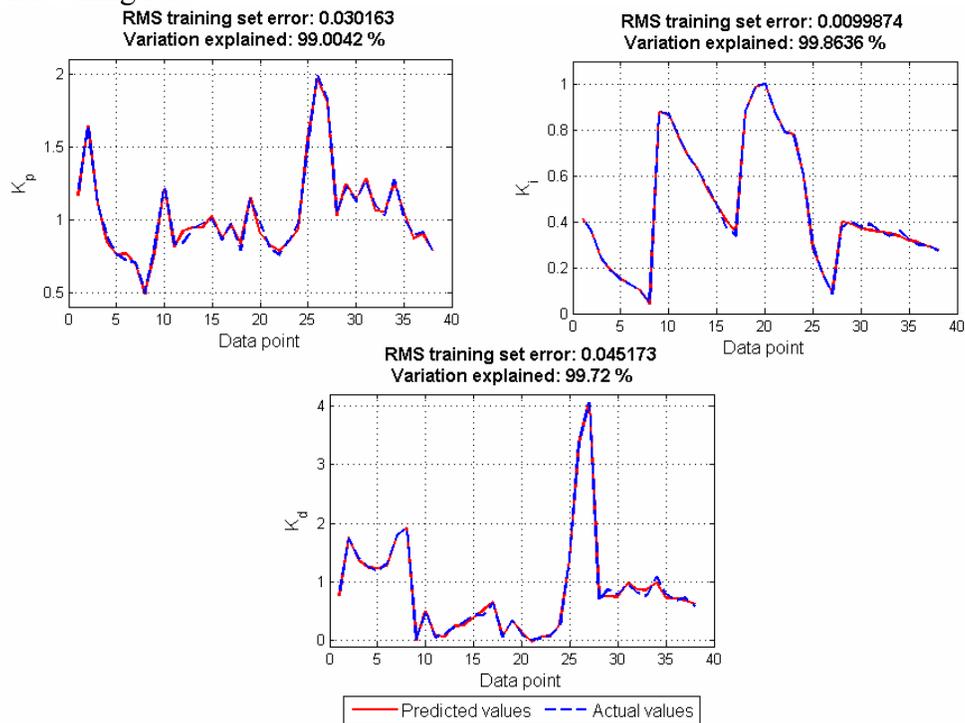

Fig. 14. Accuracies of the different PID controller parameters with the best found multi-gene tuning rule.

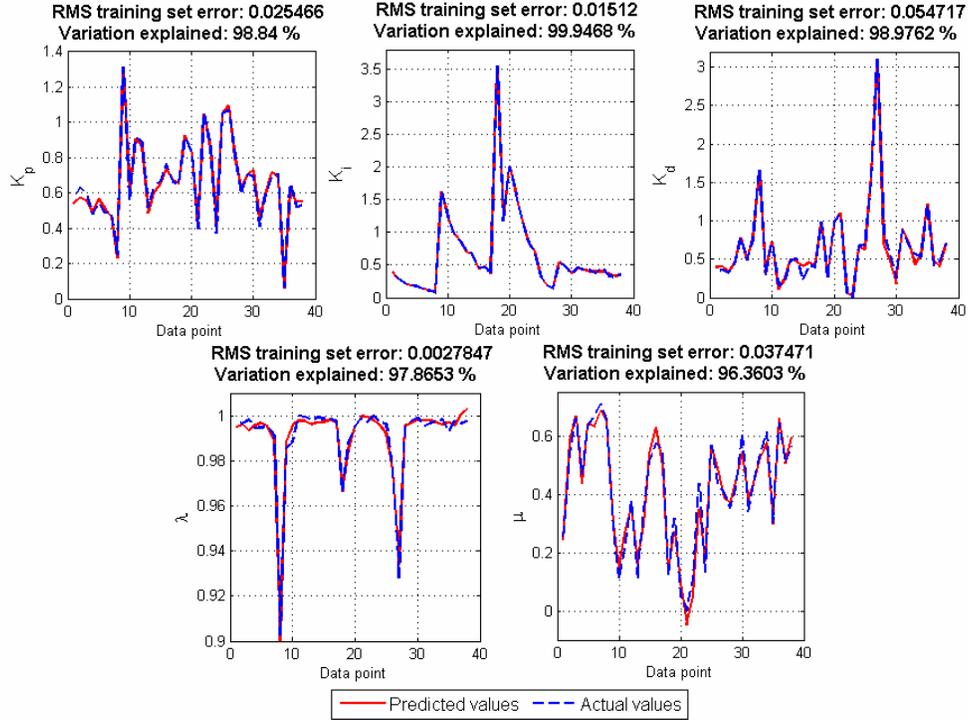

Fig. 15. Accuracies of the different $PI^{\lambda}D^{\mu}$ controller parameters with the best found multi-gene tuning rule.

PID/FOPID tuning rules, corresponding to the single and multi-gene symbolic regression approach have been reported here. Equation (13) shows the expressions for the optimal PID controller parameters obtained by the single-gene GP based symbolic regression method. It is worth mentioning that the accuracy of the rules lies in accurate reduction of higher order models in SOPTD templates with the proposed approach.

$$K_p = \frac{1}{K}\left[1.4 + 0.09685\left(\begin{array}{l}\sqrt{\left|\dfrac{\tau_{\max}}{\cos(\tau_{\min})}\right|} + \tanh\left(-\tau_{\max}^2 + \dfrac{\tau_{\min}}{\tau_{\max}}\right) + \cos\left(\dfrac{L\tau_{\max}}{\tau_{\min}}\right) + \sqrt[4]{\tau_{\max}} - \dfrac{L\tau_{\max}}{\tau_{\min}} \\ -\sin\left(1.6\times 10^{-6}\tau_{\max}^2\left(1250L + 2117\right)\left(500\left(\sqrt{\dfrac{L}{\tau_{\max}}} - \dfrac{L}{\tau_{\min}}\right) + 1877\right)\right) \\ + \tanh\left(-L + \tau_{\min}\right) - \sin\left(\tau_{\max}\right) - \dfrac{6.483756}{\tau_{\max}}\end{array}\right)\right]$$

$$K_i = \frac{1}{K}\left[1.003 - 0.2452\sqrt{\left|4\ln(\tau_{\max} + L) + 2\tanh(\tau_{\min}) + 3\tanh(L) + \tanh(\tau_{\max}) - 0.8913\right|}\right]$$

$$K_d = \frac{1}{K}\left[-1.024 + 0.539\left(\frac{\sqrt{\left|-\left(\frac{L}{\tau_{min}} + \tau_{min}^2 - \frac{541}{500}\right)\ln(\cos(\tau_{min}))(-1.031 + \cos(\tau_{min}))\right|}}{+\sqrt{\left|\frac{L^2}{\tau_{max}^{5/2}} + \cos(L) + \tau_{max}\right|} + \ln\left(\tanh\left(\frac{2.8546\tau_{min}}{L}\right) + \cos(L) + \frac{L}{\tau_{max}}\right)}\right)\right]$$

(13)

In the reported tuning rules, the division operators are expressed as protected divide, i.e., if a division by zero occurs, the term is set as zero and the other terms in the expression are evaluated to give the controller parameter values. The natural log is also defined likewise, i.e. if it does not exist or becomes undefined for certain parameter values then it is taken as zero.

Now, the best multi-gene GP based PID tuning rules are given by (14):

$$K_p = \frac{1}{K}\left[\begin{array}{l} 1.468 + 0.3362\left(\sin\left(\frac{L}{\tau_{max}}\right) - \frac{L}{\tau_{min}} - \sqrt{\left|\cos\left(\frac{\frac{L}{\tau_{min}}}{\cos(L) - \frac{L}{\tau_{max}}}\right)\right|}\right) \\ 0.1138\left(2\ln\tau_{max} - L - \frac{L}{\tau_{min}} - \sqrt{\left|\cos\left(\frac{\tau_{max}^2}{\cos(\tau_{max})}\right)\right|} - \sqrt{\left|\cos\left(\frac{1}{\tau_{min}\cos(L)}\right)\right|}\right) \\ +0.4052\left(\sqrt{\left|\cos\left(\frac{\cos(\tau_{min})}{\sin\left(\frac{L}{\tau_{min}}\right) - \frac{L}{\tau_{min}}}\right)\right|} + \cos(\cos(\sin(L)))\right) \\ -0.4627\left(\sqrt{\left|\cos\left(\frac{\tanh(L/\tau_{min})}{L}\right)\right|} + \cos(2\ln\tau_{max})\right) + 0.232\sqrt{\left|\ln(\cos(\tau_{max}))\right|} \\ +0.414\sqrt[4]{\left|\cos\left(\frac{\frac{L}{\tau_{max}} - \cos\left(\frac{L}{\tau_{min}}\right)}{\sin\left(\frac{L}{\tau_{min}}\right) - \frac{L}{\tau_{min}}}\right)\right|} - 0.3325\sqrt[4]{\left|\cos\left(\frac{\cos(L) - \frac{L}{\tau_{max}}}{\cos(L)}\right)\right|} \end{array}\right]$$

$$K_i = \frac{1}{K} \begin{bmatrix} 1.426 - 0.1283\left(\dfrac{L}{\tau_{\min}\ln\left(\tau_{\min}+\dfrac{\tau_{\max}}{\tau_{\min}}\right)}\right) - 0.0872\tanh\left(\sin\left(\ln(\tau_{\min})+51.86+\dfrac{\cos(\tau_{\min})}{L+\dfrac{L}{\tau_{\min}}}\right)\right) \\ -0.0446\tanh\left(\sin\left(\dfrac{\tau_{\max}}{\tau_{\min}}+\tau_{\max}+51.86+\dfrac{\cos(\ln(\tau_{\min}))}{5.806\tanh(\tau_{\max})}\right)\right) - 0.1572\ln(\tau_{\min}) \\ -1.268\tanh(\tanh(\tau_{\max})) - 0.0437\ln(\sin(\tau_{\max})) - 0.003763\left(\dfrac{\tau_{\max}}{\tau_{\min}}+\tau_{\max}+\dfrac{L}{\tau_{\min}\cos(\tau_{\min})}\right) \\ +0.0052\left(\tanh\left(\tau_{\min}+\dfrac{L}{\tau_{\min}\cos(\tau_{\min})}\right) + \dfrac{L}{\tau_{\min}\cos(\tau_{\min})} + \ln\left(\cos\left(\dfrac{L}{\tau_{\min}}\right)\right)\right) \end{bmatrix}$$

$$K_d = \frac{1}{K} \begin{bmatrix} 0.9524\left(\begin{array}{l}\ln\left(\dfrac{\tau_{\max}}{\tau_{\min}}(\tanh(L-7.535))\sqrt{\left|\dfrac{\tanh\left(\dfrac{\tau_{\max}}{\tau_{\min}}\right)}{\dfrac{L}{\tau_{\max}}-7.432}\right|}\right) + \dfrac{L}{\tau_{\max}} \\ +\left(\dfrac{\sqrt{L}}{\dfrac{L}{\tau_{\min}}-7.566}\right) + \left(\dfrac{0.1826}{\tau_{\max}-7.5439}\right) - \tau_{\min} - \tanh\left(\dfrac{\tau_{\max}}{\tau_{\min}} e^{\tfrac{\tau_{\max}}{\tau_{\min}}}\right)\end{array}\right) + 6.177 \\ -0.2212\left(\ln\left(\ln\left(\dfrac{\tau_{\max}}{\tau_{\min}}\right)-4.001-L\right) + \tanh\left(\cos\left(0.1247\dfrac{\tau_{\max}}{\tau_{\min}}\right)\right) - \sin\left(\sin\left(\sqrt{|\ln(L)|}\right)\right)\right) \\ -0.8712\left(\sqrt{L}+\cos\left(\sqrt{|\cos(\ln(\ln(\tau_{\min})))|}\right)\right) - 0.6186 e^{\sqrt[4]{|\cos(\ln(\ln(\tau_{\min})))|}} \\ -0.104216\dfrac{\sqrt{\left|-\tanh\left(\tau_{\min}-\dfrac{L}{\tau_{\max}}\right)\right|}}{\ln\left(\dfrac{2.391}{\tau_{\min}}+0.8314-\tau_{\max}\right)} + 2.124\ln\left(\sqrt{e^{\tau_{\min}+\ln(\tau_{\min})-\sqrt{|\cos(\tau_{\min})|}}}\right) \\ -0.163\left(\begin{array}{l}\ln\left(0.008\dfrac{\tau_{\max}}{\tau_{\min}}\left(-946+125\dfrac{L}{\tau_{\max}}\right)\cos(\ln(\tau_{\min}))\sqrt{\left|\dfrac{\tau_{\max}-7.432}{L-7.7285}\right|}\right) \\ +\dfrac{L}{\tau_{\max}} + \left(\dfrac{\sqrt{\tau_{\max}}}{\dfrac{L}{\tau_{\max}}-8.36327}\right) + \sqrt{\tau_{\max}}-\tau_{\min}\end{array}\right) \end{bmatrix}$$

(14)

The single-gene GP based optimal $PI^\lambda D^\mu$ tuning rules are also reported in (15).

$$K_p = \frac{1}{K}\left[1.188 - 0.2775\left(\frac{L}{\tau_{min}} + \tanh\left(L^2 + \frac{L}{\tau_{min}\tau_{max}} + \frac{\cos\left(\frac{\tau_{max}}{\tau_{min}}\right)}{e^{\tau_{max}}}\right) + \left(\frac{\sqrt{\left|\tanh\left(\frac{L}{\tau_{min}}\right) - \tau_{min}\right|}}{\left(\frac{\ln(L/\tau_{min})}{\tau_{max}} + 2\tau_{max}\right)\left(2(\ln(\tau_{max}))^2 + 2\frac{L}{\tau_{min}\tau_{max}^3}\right)}\right)\right)\right]$$

$$K_i = \frac{1}{K}\left[0.314 - 0.08\left(\frac{\left(\frac{\tau_{max}}{\tau_{min}}\right)^4}{\cos(L)\ln(\tau_{max}) + \tanh(\tau_{min}) + \tau_{max}}\left(\frac{\tanh(\ln(\tau_{max}))}{0.254}\right)^2 - \left(\frac{\ln(0.1851\sin(\tau_{min}))}{\tau_{max}}\right)^2\right)\right]$$

$$K_d = \frac{1}{K}\left[\begin{array}{l}0.04877 + 0.2898\cos\left(\frac{\tau_{max}}{\tau_{min}}\right) \\ +0.1449\left(\begin{array}{l}\left((\tau_{min} - 1.972)^2\sin^2(\sin(\tau_{max})) + \sqrt{\left|\sin\left(\frac{4.86129}{L}\right)\right|}\right) \\ -\sin(\tau_{max}) + \sqrt{\left|\sin\left(\frac{\tau_{max}}{\tau_{min}}\right)\right|} + \sqrt{\left|\sin\left(-\frac{9.56649}{\tau_{min}}\right)\right|} + \ln\left(\sin\left(\frac{\tau_{max}}{\tau_{min}}\right)\right) \\ +\sqrt{\left|\ln\left(\frac{\tau_{max}}{\tau_{min}}\right)\right|} + \sqrt{\left|\sin\left(-\frac{9.61668}{\tau_{min}}\right)\right|} - \cos\left(\frac{L}{\tau_{max}}\right) + \cos\left(4.735\frac{\tau_{max}}{\tau_{min}}\right)\end{array}\right)\end{array}\right]$$

$$\lambda = 0.9974 - 0.002605\sqrt{\tau_{max}L}(\tau_{max} - \tanh(\tau_{min}))$$

$$\mu = 2.0205 + 1.708\left[\begin{array}{l}\tanh(\tanh(\tanh(L))) - \cos\left(\tanh\left(\frac{\tau_{max}}{\tau_{min}}\right)\right) - \cos\left(\cos\left(\tanh\left(\frac{L\tau_{max}}{\tau_{min}}\right)\right)\right) \\ -\cos\left(\tanh\left(L + \frac{L}{\tau_{max}} + \frac{\tau_{max}^2}{\tau_{min}^2}\right)\right) + \cos\left(\cos\left(\frac{L\tau_{min}}{\tau_{max}^2 e^{\frac{\tau_{max}}{\tau_{min}}}}\right)\right)\end{array}\right]$$

(15)

The best multi-gene $PI^\lambda D^\mu$ controller tuning rules are given in (16).

$$K_p = \frac{1}{K} \begin{bmatrix} -5.94177 \tanh\left(\sqrt[4]{\left|\ln\left(7.964 + \frac{L}{\tau_{min}} - \frac{\tau_{max}}{\tau_{min}}\right)\right|}\right) - 1.146\sqrt{\left|\sin\left(\sin\left(\sin\left(\sin\left(\frac{\tau_{max}^2}{\tau_{min}^2}\right)\right)\right)\right)\right|} \\ +0.04561\begin{pmatrix} \cos\left(\left(-8049 - 1000\frac{L}{\tau_{min}} + 1000\frac{\tau_{max}^2}{\tau_{min}^2}\right)\left(0.001\frac{L\tau_{max}^2}{\tau_{min}^3}\right)\right) \\ +\cos\left(8.757\frac{L\tau_{max}^2}{\tau_{min}^3}\right) + \sqrt{\left|\frac{\tau_{max}^2}{\tau_{min}^2}\tanh\left(\frac{L}{\tau_{max}}\right)\middle/\tanh\left(\frac{\tau_{max}}{\tau_{min}}\right)\right|} \end{pmatrix} + 6.5488 \\ -0.2383 \cos\left(\ln\left(\tanh\left(\frac{L}{\tau_{min}}\tanh(L) + \sin\left(\tau_{min} + \frac{\tau_{max}}{\tau_{min}}\right)\right)\right)\right) \\ +0.24655 \cos\left(\ln\left(\tanh\left(\frac{L}{\tau_{max}} + \ln\left(\frac{L^2}{\tau_{max}\tau_{min}}\right)\right)\right)\right) \\ -0.1022\sqrt{\left|\ln\left(\frac{2\tau_{max}}{\tau_{min}}\right)\frac{L\tau_{max}}{\tau_{min}^2}\frac{\sin\left(\frac{L^2}{\tau_{max}\tau_{min}}\right)}{\ln\left(\frac{L^2}{\tau_{max}\tau_{min}}\right)} + \cos\left(0.284 + \frac{\tau_{max}^2}{L} - \sqrt{\tau_{min}}\right)\right|} \\ +0.2571\left(\cos(\ln(L)) - \sin\left(\cos\left(\cos\left(e^{\frac{7.646L}{\tau_{max}}}\right)\right)\right)\right) \end{bmatrix}$$

$$K_i = \frac{1}{K}\begin{bmatrix} 1.6428 - 0.01641\begin{pmatrix} \ln\left(\ln(L) + \frac{\tau_{\max}}{\tau_{\min}} + \ln(\tau_{\min})\right) + \cos\left(\frac{\tau_{\max}^2}{\tau_{\min}^2} + \frac{\tau_{\max}}{\tau_{\min}} + \frac{e^{\frac{L}{\tau_{\max}}}}{\ln(L)}\right) \\ + \ln(\tau_{\min}) + \frac{\tau_{\max}}{\tau_{\min}} + \tanh\left(\frac{\tau_{\max}^2}{\tau_{\min}^2}\right) + \ln\left(\frac{L^2}{\tau_{\max}^2}\right) \end{pmatrix} \\ -0.02497 e^{\tanh(\tau_{\max}) + \tanh(e^{\tau_{\min}}) + \tanh(3\tau_{\max}) + \tanh\left(1.989 + \left(e^L/\tau_{\min}^2\right)\right)} \\ -0.00019\tau_{\max}^2 - 0.00009464\left(\left(\frac{\frac{L}{\tau_{\max}\tau_{\min}}}{\ln(L)}\right)^2 + \frac{\frac{\tau_{\max}^2}{\tau_{\min}^2}\sinh\left(\ln\left(\frac{L}{\tau_{\min}}\right)\right) + \frac{\tau_{\max}}{\tau_{\min}} + \frac{e^{L/\tau_{\max}}}{\ln(L)}}{\cosh\left(\ln\left(\frac{L}{\tau_{\min}}\right)\right)\cos(\ln(\tau_{\min}))}\right) \\ -0.0008462\frac{L^4}{\tau_{\max}^4} + 0.059\frac{\tau_{\max}}{\tau_{\min}} + 0.0295\left(\frac{\tau_{\max}^2}{\tau_{\min}^2} + \ln(\ln(\tau_{\min})) + \ln\left(\cos\left(e^{2\tau_{\min}}\right)\right)\right) \\ +0.02669\ln\left(\tanh\left(L + \frac{L}{\tau_{\min}\ln(L)}\right)\right)\left(\cos(\tau_{\min}) + \cos\left(e^{\frac{2\tau_{\max}}{\tau_{\min}}}\right) + \ln(\tau_{\min})\right) \\ -0.03\left(\ln\left(\tanh\left(L + \frac{\frac{L}{\tau_{\max}^2}}{\ln(L)}\right)\right) + \frac{\tau_{\max}^2}{\tau_{\min}^2}\right) + 9.464\times10^{-5}\left(\frac{e^L}{\tau_{\max}^2} - \left(\tau_{\min} + \frac{\tau_{\max}}{\tau_{\min}}\right)^2\right) \\ -0.03295\begin{pmatrix} \cos(\ln(\ln(\tau_{\min})))^2 + \tanh\left(\frac{L}{\ln(L) + \frac{\tau_{\max}^4}{\tau_{\min}^4}}\right) + \cos\left(\ln\left(\frac{L}{\tau_{\min}}\right) + \frac{\tau_{\max}}{\tau_{\min}}\right) \\ -\sin\left(-\frac{L^2}{\tau_{\max}^2} + 0.7031 + L\right) + \tau_{\min} \end{pmatrix} \end{bmatrix}$$

$$K_d = \frac{1}{K} \begin{bmatrix} 3.453 - 4.196\cos\left(\cos\left(\cos\left(\cos\left(\cos\left(\tau_{min} + \frac{\tau_{max}}{\tau_{min}}\right)\right)\right)\right)\right) \\ -0.3846\left(\sin\left(\sin\left(\sqrt{\frac{\tau_{max}}{\tau_{min}}}\right)\right) - \tau_{max}\right) + 0.1009\ln\left(\frac{\sqrt{|\cos(\tau_{max})|}}{\frac{\tau_{max}}{\tau_{min}} + L - e^{\tau_{min}}}\right) \\ +0.008964 \frac{\sin\left(\sqrt{\frac{\tau_{max}}{\tau_{min}}}\right) + \frac{L}{\tau_{max}} - \cos\left(\tau_{min} - \tanh\left(\frac{L}{\tau_{min}}\right)\right) + \tau_{min}}{\tanh\left(\tanh\left(\tau_{min} - \frac{L^2}{\tau_{min}^2}\right)\right)} \\ -0.131\cos\left(e^{2\tau_{min}}\left(\ln\left(\frac{L}{\tau_{max}}\right) - \frac{L}{\tau_{min}} + L\right) + \cos\left(-2.021 + \frac{\tau_{max}}{\tau_{min}} - \frac{\tau_{max}^2}{\tau_{min}^2}\right)\right) \\ +0.08997\cos\left(e^{2\left(\tau_{min} + \frac{\tau_{max}}{\tau_{min}}\right)}\left(-\tau_{min} - \frac{2\tau_{max}}{\tau_{min}} + \tau_{max}\right) + e^{2\tau_{min}}\left(-\ln\left(\frac{L}{\tau_{max}}\right) + \frac{L}{\tau_{min}} - L\right)\right) \\ +0.1828\cos\left(e^{2\tau_{min}}\left(2\tau_{min} - \frac{L}{\tau_{min}} + L\right) + \cos(\tau_{min})\right) \end{bmatrix}$$

$$\lambda = -0.0001\frac{\tau_{max}}{L} + 0.261\left(\ln\left(\cos\left(\tanh\left(\tau_{max} + 6.405\right)\right)\right)\right)^2 + 0.01138\frac{L}{\tau_{min}} + 0.00264\ln\left(\frac{\tau_{max} - L}{\tau_{min}}\right)$$
$$- 0.0001788\left(\left(e^{\tanh(\tau_{min})}(L + \tau_{max})^2\right) + \cos^2\left(\frac{L^2}{\tau_{min}^2}\right)\right) + 0.004568\sqrt{\left|e^{\tanh(L/\tau_{min})}\right|} + 0.88968$$

$$\mu = 0.876643\tanh(\tanh(L)) + 0.0055\left(\left(\sin\left(\frac{\left(\frac{\tau_{max}/\tau_{min}}{\tanh(L/\tau_{min})}\right)}{\ln(\tanh(L))}\right) + \frac{\cos\left(e^{L/\tau_{min}}\right)}{\cos\left(L - \frac{L}{\tau_{min}}\right)}\right) \Big/ \ln(L/\tau_{min})\right)$$
$$- 0.06738\ln\left(\ln\left(\frac{L}{\tau_{min}}\right) + \cos\left(\frac{\tau_{max}}{\tau_{min}}\right)\right) - 0.2712\cos\left(\frac{\tau_{max} - L}{\tau_{min}}\right) + 0.01116 e^{e^{\sin\left(\frac{0.8896}{2\ln(\tau_{max}/\tau_{min})}\right)}}$$
$$+ 0.04845\ln\left(\sin\left(\sin\left(\frac{\frac{L}{\tau_{max}} + \frac{L}{\tau_{min}}}{\ln(\cos(L))}\right)\right)\right) - 0.0505\ln\left(\sin\left(\sin\left(\frac{\frac{\tau_{max}/\tau_{min}}{\tanh(L/\tau_{min})}}{\ln(\tanh(L))}\right)\right)\right) + 0.06646$$

(16)

At a glance, these tuning rules may seem to be very complex but they perfectly maps the global optimization (GA) based PID/FOPID controller parameters corresponding to each reduced SOPTD process. O' Dwyer [1] has given many complicated analytical PID tuning rules especially for SOPTD systems with minimum integral error index. The present study reports the analytical tuning rules for PID/FOPID controllers with minimum error index as well as controller effort for wide variety of higher order process including repeated pole and non-minimum phase processes.

### *3.4. Visualization of the optimal PID/FOPID tuning rules:*
### *3.4.1. Optimal FOPID tuning rules:*

The best FOPID rules which also have the highest complexity, for the single and multi gene GP cases are compared with respect to variation in the two time constants and delay. As is evident from the Fig. 16-20, the variations for the single gene cases are much smoother than the multi gene cases. This is due to the fact that the tuning rules, evolved with the multi-gene cases are much more complex than their single gene counterparts and hence can account for the non-linear interrelationship of the reduced process parameters with the tuned controller parameters in a better way. Also in many cases, there is unevenness in the parameter landscape which is clustered at a specific range of $\tau_{min}$ and $\tau_{max}$ while at other values the curve is relatively flat. This is due to the fact that the $\tau_{min}$ and $\tau_{max}$ of the test-bench plants lie in that range and hence the tuning rules have been effective in mapping the parameter values in those specific regions. Outside these regions the tuning rules can approximately model the variation and are relatively less accurate.

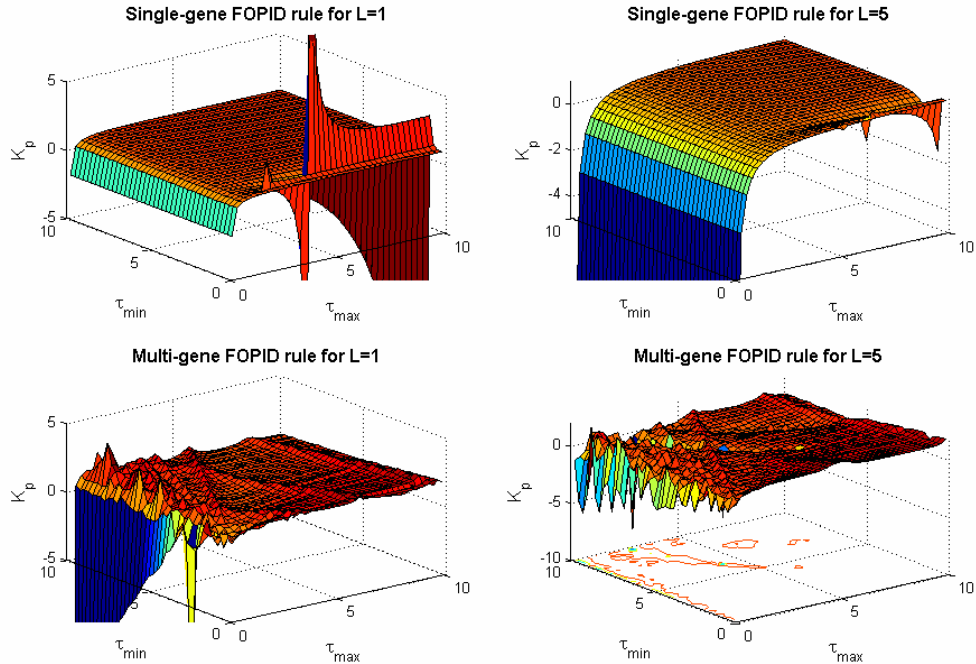

Fig. 16. 3D visualization of proportional gain ($K_p$) of FOPID controller.

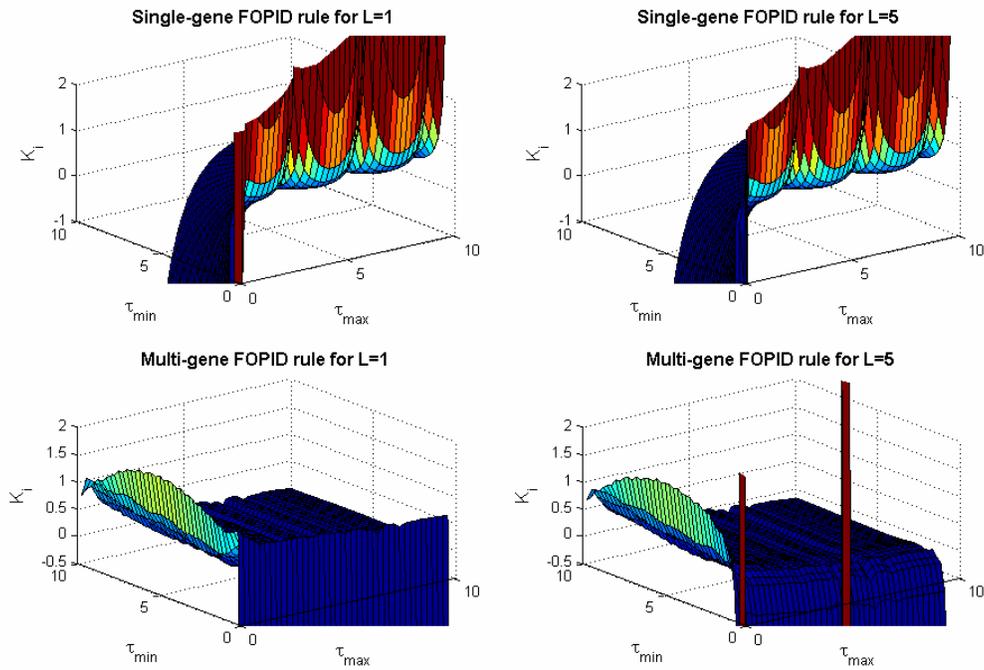

Fig. 17. 3D visualization of integral gain ($K_i$) of FOPID controller.

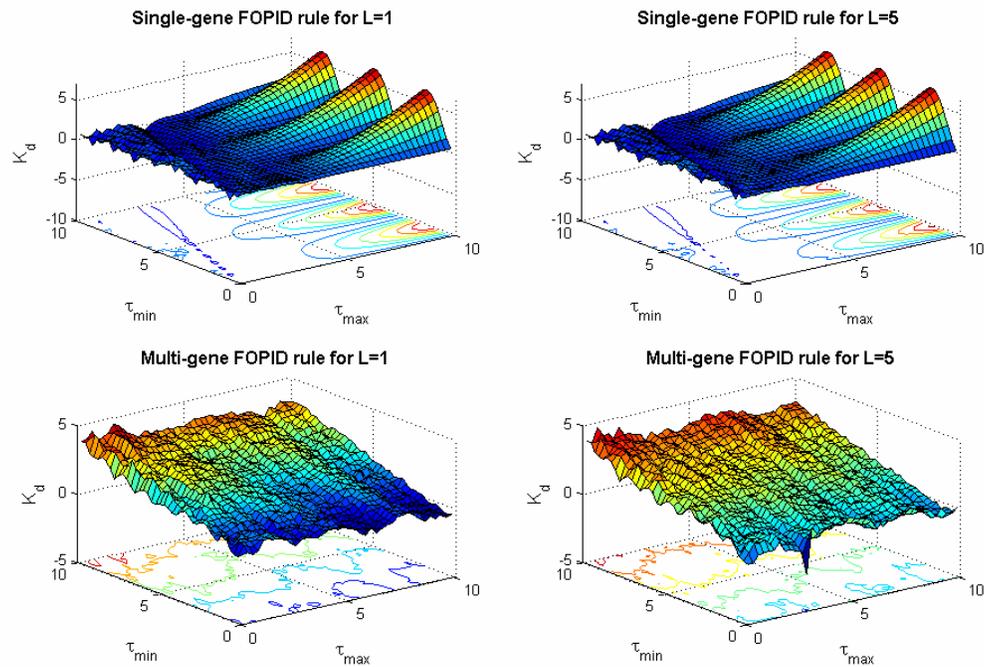

Fig. 18. 3D visualization of derivative gain ($K_d$) of FOPID controller.

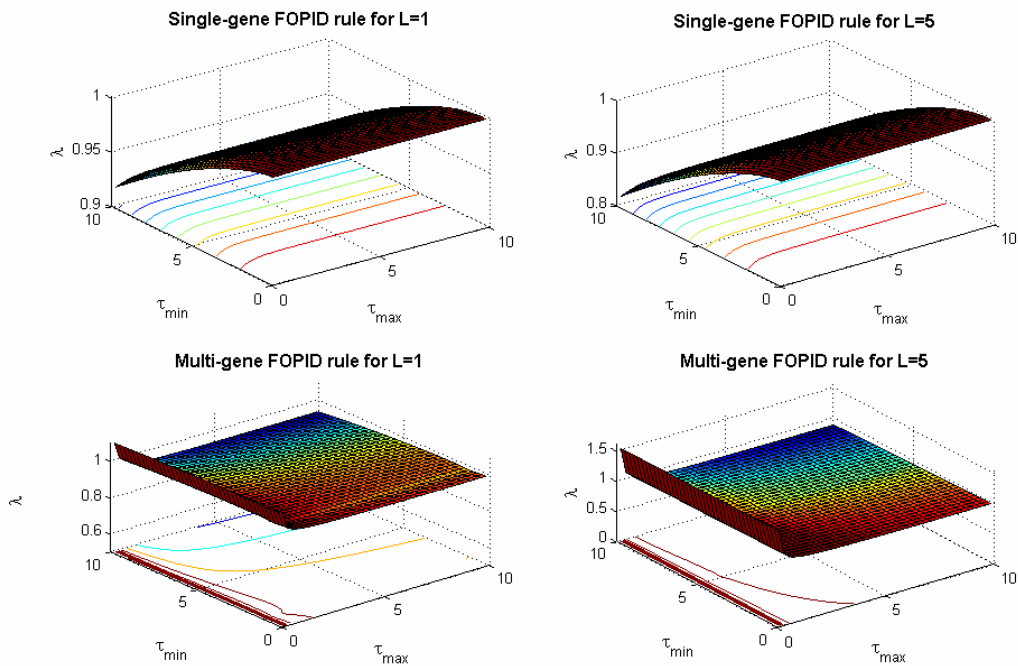

Fig. 19. 3D visualization of integral order (λ) of FOPID controller.

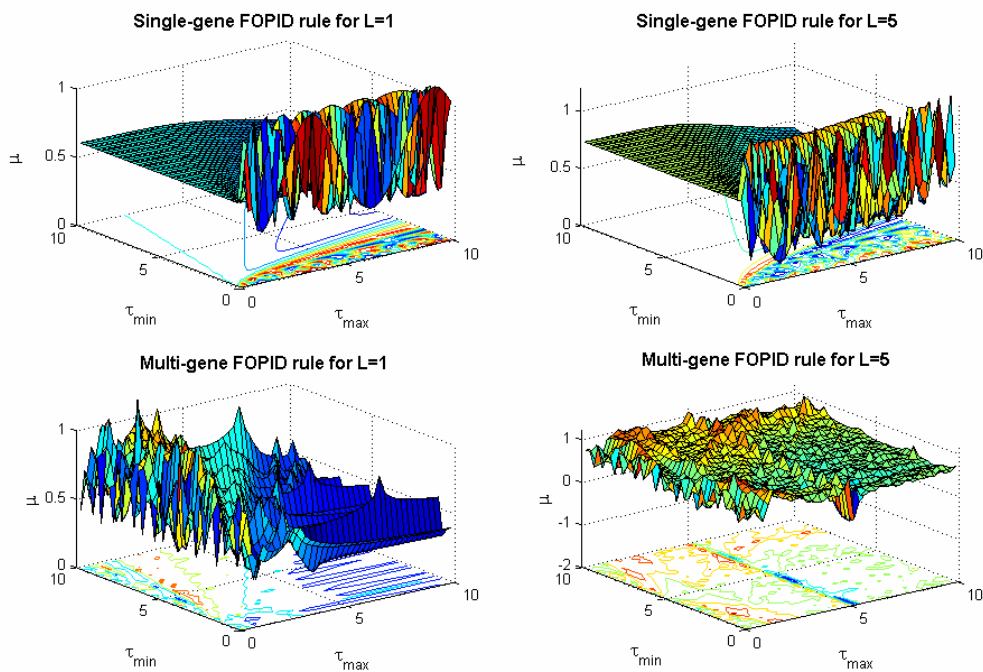

Fig. 20. 3D visualization of derivative order (μ) of FOPID controller.

*3.4.2. Optimal PID tuning rules:*

The best single/multi-gene PID tuning rules are shown in Fig. 21-23 with variation in delay and time constants.

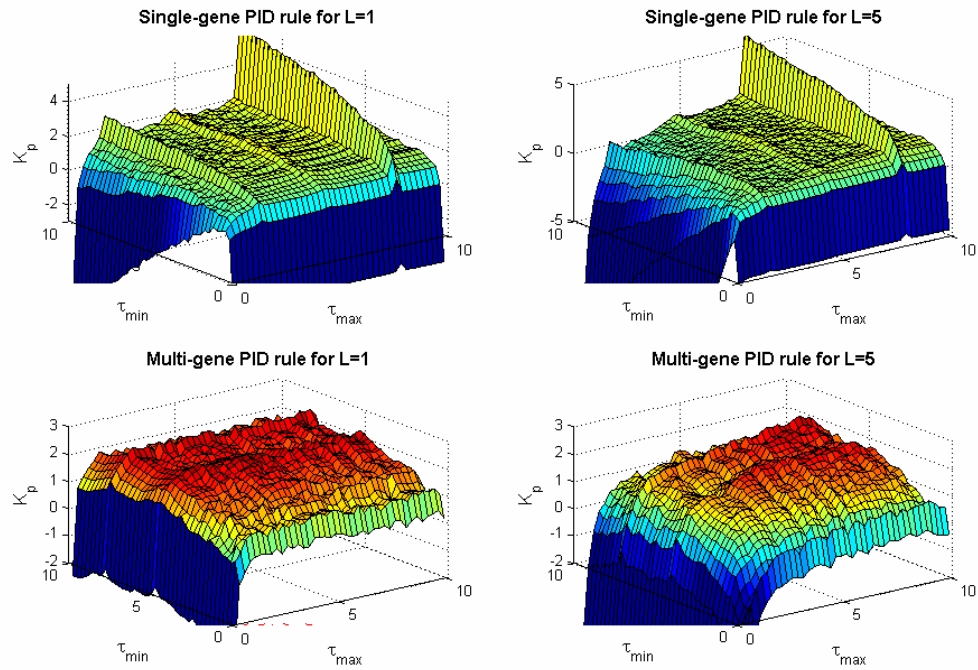

Fig. 21. 3D visualization of proportional gain ($K_p$) of PID controller.

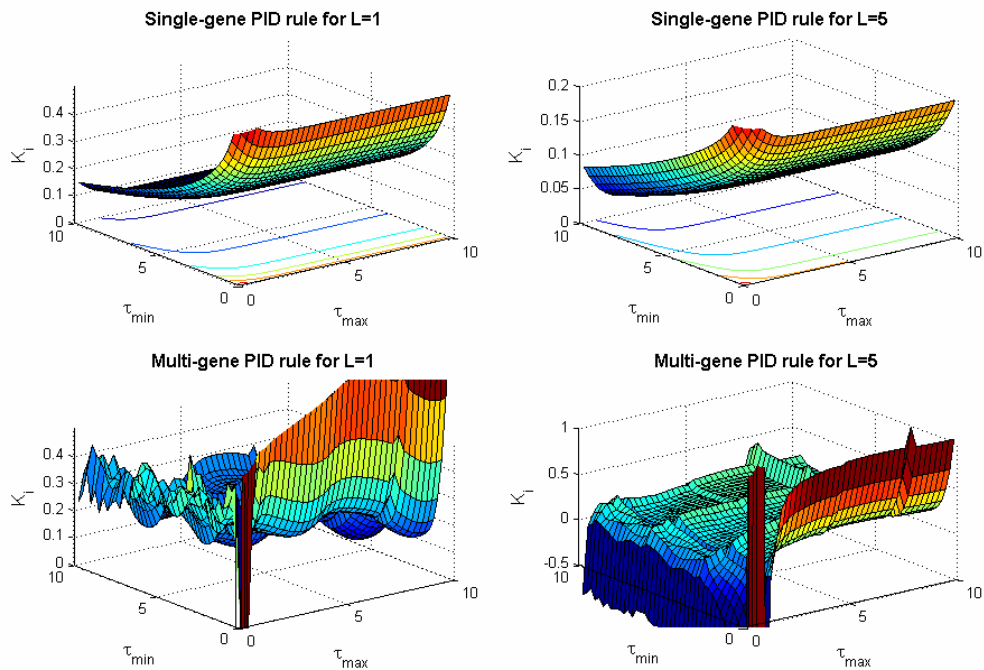

Fig. 22. 3D visualization of integral gain ($K_i$) of PID controller.

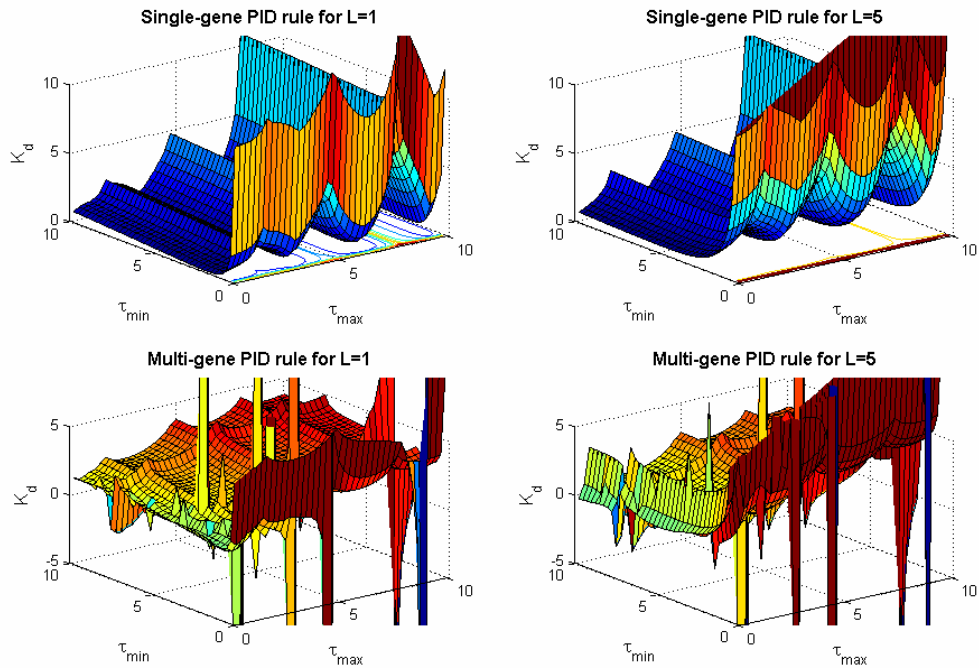

Fig. 23. 3D visualization of derivative gain ($K_d$) of PID controller.

Table 5: Optimal controller parameters with the GP based tuning rules

| Process | Controller | Type of Rule | $K_p$ | $K_i$ | $K_d$ | $\lambda$ | $\mu$ |
|---|---|---|---|---|---|---|---|
| $P_1$ (n=8) | FOPID | single-gene GP | 0.4156 | 0.1517 | 0.6362 | 0.9872 | 0.6539 |
| | | multi-gene GP | 0.4605 | 0.1279 | 0.4842 | 0.9954 | 0.6508 |
| | | GA based optimized | 0.4895 | 0.129 | 0.4882 | 0.9952 | 0.669 |
| | PID | single-gene GP | 0.8197 | 0.1439 | 1.3342 | - | - |
| | | multi-gene GP | 0.7619 | 0.1246 | 1.2752 | - | - |
| | | GA based optimized | 0.717 | 0.126 | 1.3042 | - | - |
| $P_2$ ($\alpha$=0.6) | FOPID | single-gene GP | 0.7092 | 0.6648 | 0.6111 | 0.9971 | 0.3314 |
| | | multi-gene GP | 0.6027 | 0.6612 | 0.4647 | 0.9969 | 0.2957 |
| | | GA based optimized | 0.635 | 0.6679 | 0.5089 | 0.9991 | 0.3441 |
| | PID | single-gene GP | 0.8863 | 0.5405 | 0.3038 | - | - |
| | | multi-gene GP | 0.9443 | 0.5413 | 0.2673 | - | - |
| | | GA based optimized | 0.973 | 0.5542 | 0.3119 | - | - |
| $P_3$ (T=5) | FOPID | single-gene GP | 0.9506 | 0.1358 | 1.4809 | 0.9738 | 0.4479 |
| | | multi-gene GP | 1.0933 | 0.1935 | 1.5977 | 0.976 | 0.4898 |
| | | GA based optimized | 1.0698 | 0.1922 | 1.5295 | 0.977 | 0.4368 |
| | PID | single-gene GP | 1.9955 | 0.1737 | 3.3979 | - | - |
| | | multi-gene GP | 1.9684 | 0.1761 | 3.3905 | - | - |
| | | GA based optimized | 1.993 | 0.1708 | 3.3888 | - | - |
| $P_4$ ($\alpha$=0.4) | FOPID | single-gene GP | 0.6254 | 0.3948 | 0.6054 | 0.996 | 0.5038 |
| | | multi-gene GP | 0.4506 | 0.4424 | 0.8921 | 0.9979 | 0.3751 |
| | | GA based optimized | 0.4129 | 0.4615 | 0.8846 | 0.9952 | 0.337 |
| | PID | single-gene GP | 1.1572 | 0.3569 | 0.759 | - | - |
| | | multi-gene GP | 1.2803 | 0.3631 | 0.9782 | - | - |
| | | GA based optimized | 1.2648 | 0.3884 | 0.9708 | - | - |

These 3-dimensional plots representing PID/FOPID parameters with variation in time constants and delay are especially important as a guideline for manual variation in controller knobs so as to maintain good set-point tracking performance with the requirement of low control signal.

### *3.5. Performance of the analytical tuning rules:*

Four representative processes have been chosen from the four different classes of higher order processes (7)-(10) to validate the PID/FOPID tuning formula obtained by GP. Also, the GA based optimum control performances are compared with the rule based PID/FOPID controller, to show the wide applicability of such rules in process controls. Table 5 shows the computed PID/FOPID controller parameters for four representative processes among the test-bench. The simulated time response and control signals have been shown in Fig. 24-27. The figures indicate that the set-point tracking, load disturbance rejection and control signals of the multi gene GP rules are closer to those obtained by GA based methods and are better than their single gene GP rule counterparts. Hence, the PID and FOPID tuning rules given by (14) and (16) can be used for wide variety of processes given by (7)-(10). The effectiveness of such GP based optimal PID/FOPID tuning rule extraction can be viewed like combining the capability of set-point tracking, load disturbance rejection and small control signals in a single rule to handle a wide range of stable higher order processes, commonly encountered in process control industries [39].

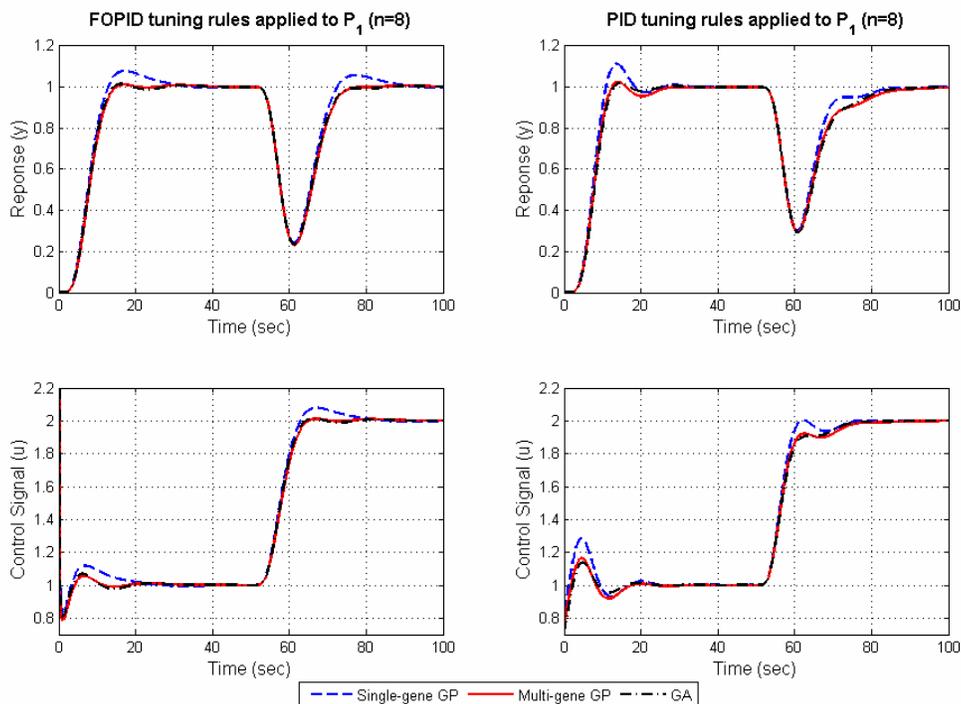

Fig. 24. Performance of the optimum PID/FOPID tuning rules for plant $P_1$.

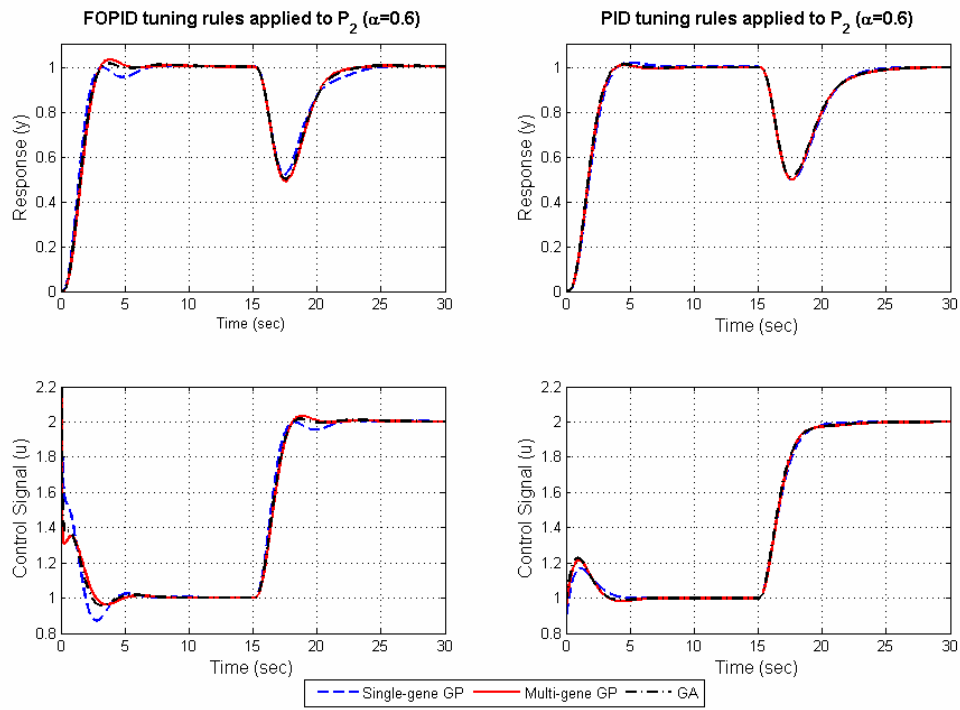

Fig. 25. Performance of the optimum PID/FOPID tuning rules for plant $P_2$.

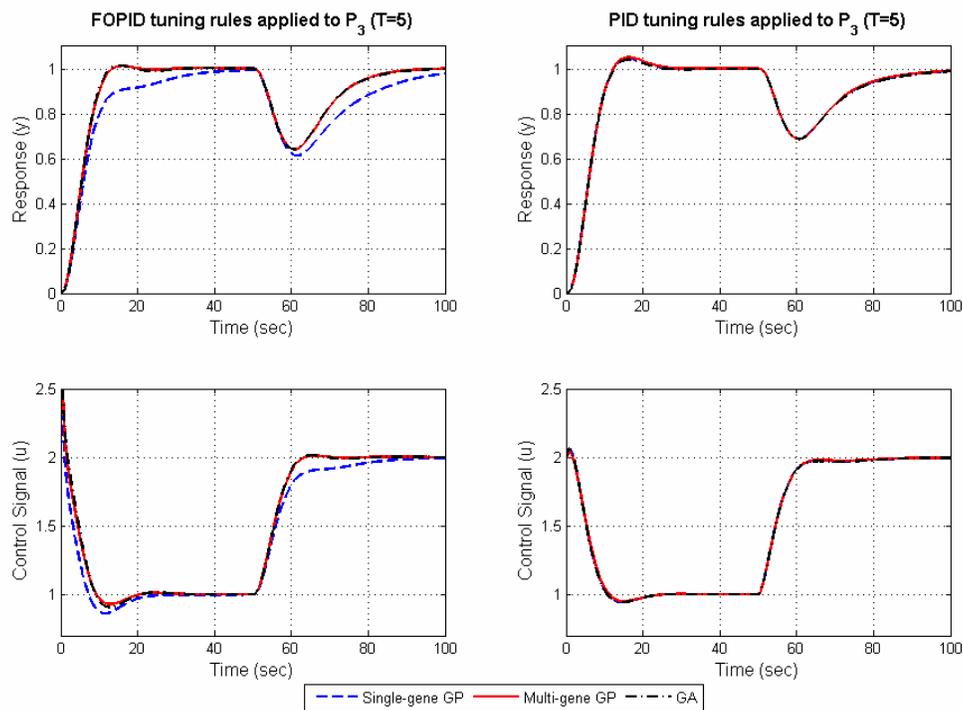

Fig. 26. Performance of the optimum PID/FOPID tuning rules for plant $P_3$.

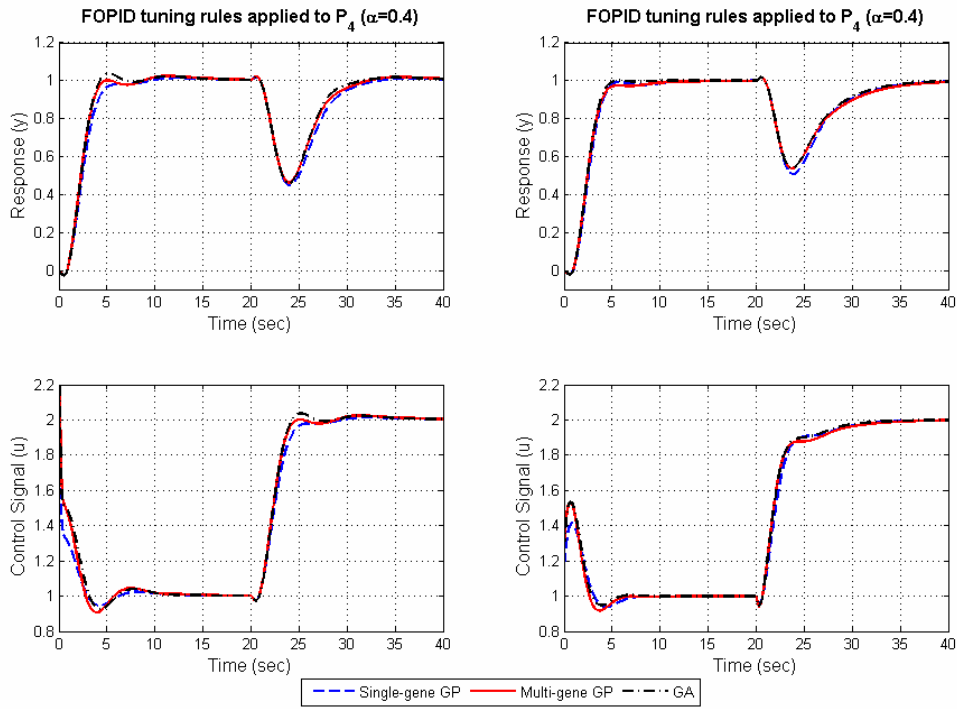

Fig. 27. Performance of the optimum PID/FOPID tuning rules for plant $P_4$.

### *3.6. Effect of plant perturbation on the tuning rules:*

In practice, higher order process models can be inaccurately reduced or the initial higher order models can be erroneously estimated leading to plant parameter uncertainty. A good tuning rule should be capable of taking these uncertainties into account while also maintaining the control performance. In order to test the inherent robustness of these optimal PID/FOPID tuning rules simulations are carried out for variation in process dc-gain ($K$), maximum time constant ($\tau_{\max} = T_0$) and delay ($L = L_0$) of the controlled process. Recent literatures report few interesting results on fractional order controller for handling plant uncertainties like dc-gain [24], [52]-[53], time constant [54]-[56] and time delay [57]-[58] and improvement in control performance has also been shown. Similar to the mentioned literatures ±10% variation in dc-gain [52], ±20% variation in dominant time constant [54] and ±50% variation in time-delay [57] has been done with the controller parameters reported in Table 5. It can be seen from Fig. 28-29 that the tuning rules gives sufficient parametric robustness to the PID and FOPID controllers for maintaining satisfactory control performance although there was no explicit consideration of plant uncertainty while developing these rules. Fig. 28-29 also indicate that the multi-gene tuning rules are more robust than the single gene tuning rules and can give good set point tracking and load disturbance rejection even under plant uncertainty.

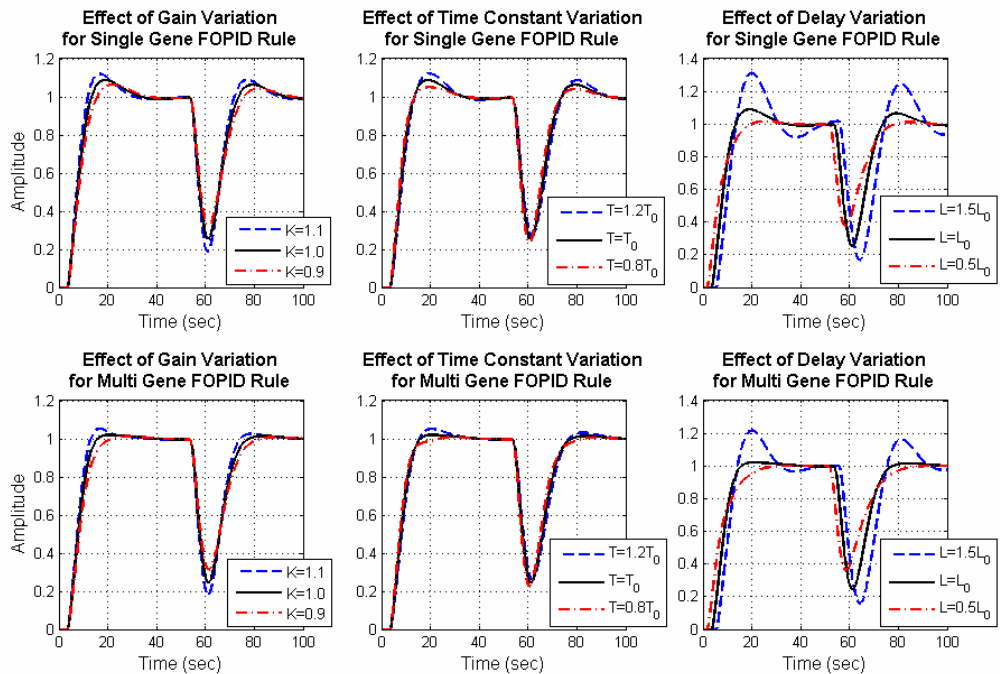

Fig. 28. Time responses for the rule based FOPID controllers with plant uncertainty.

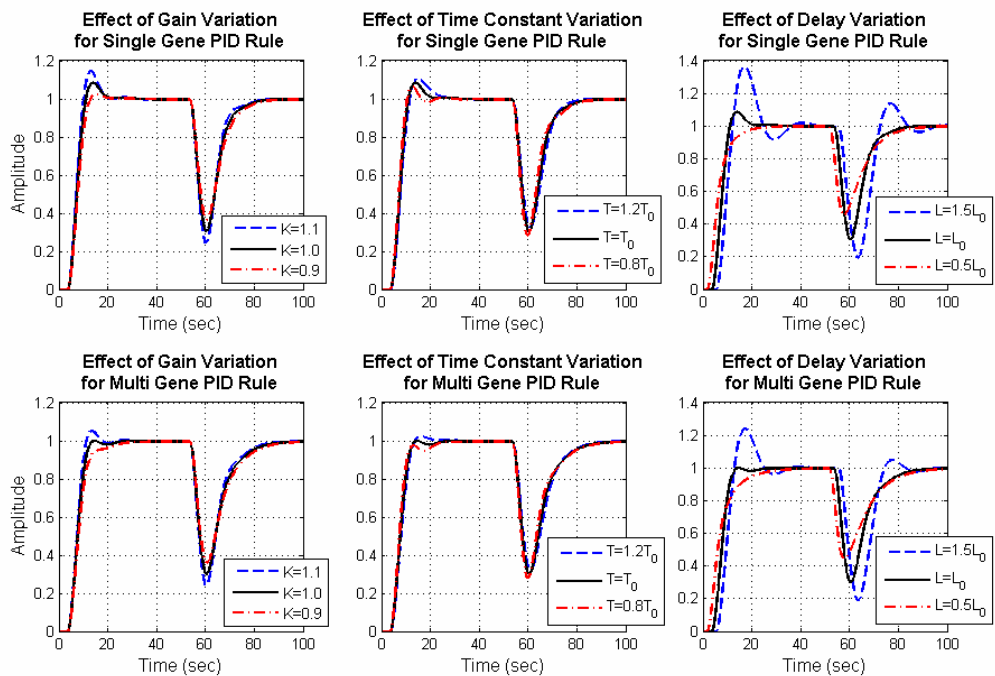

Fig. 29. Time responses for the rule based PID controllers with plant uncertainty.

## 4. Conclusion:

An improved Nyquist based sub-optimal model reduction technique using Genetic Algorithm has been proposed in this paper which outperforms the existing $H_2$ norm based model reduction technique. Four different class of higher order processes are modeled in FOPTD and SOPTD template using the proposed technique. GA is also employed to tune optimal PID and $PI^\lambda D^\mu$ controllers while minimizing an objective function, comprising of error index and controller effort. With the GA based sub-optimal SOPTD model parameters and GA based PID and FOPID controllers optimal tuning rules are extracted via a symbolic regression technique known as Genetic Programming. The rules are in the form of analytical expressions and hence are valuable to process control engineers due to ease of calculation and online implementation. These rules are also very useful in real time automation as they can be embedded in practical hardware to control a non-linear or time varying plant which can be identified online and reduced to SOPTD template. Tuning rules with best fit and highest complexity have been reported in this paper for single and multi-gene GP. Multi-gene PID/FOPID rules gives better control performance and robustness as they mimic the GA based results more accurately. The performance of the single/multi-gene optimum tuning rules is demonstrated vis-à-vis the original GA based controller parameters, indicating nominal deterioration in the closed loop response of the overall control system. Three dimensional plots of PID/FOPID controller parameters (gain and orders) are shown as a guideline for process operators. Robustness of the rules against plant dc-gain, dominant time-constant and delay variation have also been demonstrated. Future scope work may include, similar tuning rule generation with frequency domain controller tuning methods.

**Acknowledgement:**

This work has been supported by the Department of Science and Technology (DST), Govt. of India under the PURSE programme.